\documentclass[preprint,3p,12pt]{elsarticle}

\usepackage{amssymb}
\usepackage{amsmath}
\usepackage{comment}
\usepackage{url}

\usepackage{acronym}
\newacro{PIV}{Particle Image Velocimetry}
\newacro{FOV}{Field of View}
\newacro{POD}{Proper Orthogonal Decomposition}
\newacro{EPOD}{Extended Proper Orthogonal Decomposition}
\newacro{SVD}{Singular Value Decomposition}
\newacro{MSE}{Mean Square Error}
\newacro{MAE}{Mean Absolute Error}
\newacro{LSM}{Least Squares Method}
\newacro{RNN}{Recurrent Neural Network}
\newacro{DNS}{Direct Numerical Simulation}
\newacro{JHTDB}{Johns Hopkins Turbulence Databases}
\newacro{2D}{Two-Dimensional}
\newacro{3D}{Three-Dimensional}
\newacro{UC3M}{Universidad Carlos III de Madrid}
\newacro{AMIC}{Advection-based Multiframe Iterative Correction}
\newacro{SMART}{Simultaneous Multiplicative Algebraic Reconstruction Technique}
\newacro{cSMART}{camera-Simultaneous Multiplicative Algebraic Reconstruction Technique}
\newacro{PTV}{Particle Tracking Velocimetry}
\newacro{LOR}{Low Order Reconstruction}
\newacro{GT}{Ground Truth}
\newacro{RMS}{Root-Mean-Square}
\newacro{PDF}{probability density function}
\newacro{LSE}{Linear Stochastic Estimation}
\newacro{ML}{Machine Learning}
\newacro{AMIC}{Advection-based Multiframe Iterative Correction }

\journal{XXX}

\begin{document}

\begin{frontmatter}



\title{Data-efficient semi-supervised learning for flow estimation using unlabelled probe data}




\author[1,2]{Junwei Chen}
\ead{chenjunwei@pku.edu.cn}
\author[2]{Marco Raiola}
\ead{mraiola@ing.uc3m.es}
\author[2]{Stefano Discetti}
\ead{stefano.discetti@uc3m.es}

\affiliation[1]{organization={Peking University, State Key Laboratory for Turbulence and Complex Systems}, addressline={No.5 Yiheyuan Road, Haidian District}, city={Beijing}, postcode={100871}, state={}, country={China}}
\affiliation[2]{organization={Universidad Carlos III de Madrid, Department of Aerospace Engineering}, addressline={Avda. Universidad 30}, city={Leganés}, postcode={28911}, country={Spain}}

\begin{abstract}
Estimating time-resolved velocity and pressure fields from \ac{PIV} remains challenging due to its limited temporal resolution in many applications. Data-driven approaches that combine snapshot \ac{PIV} with high-frequency probe data have shown great promise in reconstructing the flow dynamics for advection-dominated flows; however, they typically exploit only the probe measurements directly synchronized with the \ac{PIV} frames, leaving a large volume of probe-only data acquired between snapshots unused. In this work, we propose a framework that enriches the original \ac{PIV} training dataset by time-marching a simple advection model and then exploits unlabelled probe data through a semi-supervised learning strategy. Two neural networks are trained to predict the temporal coefficients of \ac{POD} modes of the flow fields, and their temporal derivatives, respectively. Unlabelled probe samples are leveraged to enforce temporal consistency and expand the coverage of flow scenarios beyond those captured by snapshot \ac{PIV}, which is crucial for obtaining physically consistent temporal gradients required for pressure field reconstruction. A least-squares regularization step is further employed to reconcile the predictions and enforce consistency between temporal coefficients and their derivatives. The proposed approach is validated on both synthetic turbulent channel flow data and experimental \ac{PIV} measurements of an airfoil wake. Results demonstrate that incorporating unlabelled probe data significantly improves the accuracy and temporal smoothness of velocity reconstruction, leading to more reliable pressure estimation via the Navier–Stokes equations, without increasing the experimental cost.
\end{abstract}



\begin{keyword}
Proper Orthogonal Decomposition \sep Artificial Intelligence \sep Machine Learning \sep Particle Image Velocimetry



\end{keyword}

\end{frontmatter}



\acresetall
\section{Introduction}
\label{sec: intro}


\ac{PIV} offers spatially-resolved velocity information and has become a standard tool for experimental fluid mechanics. However, achieving sufficiently high temporal resolution with \ac{PIV} requires high-repetition-rate lasers and cameras, which significantly increase experimental complexity and cost \cite{beresh2021time}. Many experiments thus rely on snapshot \ac{PIV} (i.e. with frame rate significantly smaller than the characteristic frequencies of the flow), which provides limited information on the dynamics and challenges the estimation of quantities such as pressure fields.

To address this limitation, data-driven methods have been developed to infer time-resolved flow information by combining low-frame-rate \ac{PIV} snapshots with high-frequency point measurements, such as pressure transducers or hot-wire probes. Classical techniques including \ac{LSE} \cite{adrian1988stochastic} and \ac{EPOD} \cite{boree2003extended} exploit statistical correlations between sparse probe signals and spatial velocity fields. These approaches have been successfully applied in a range of flows to reconstruct unsteady velocity fields and, in some cases, estimate pressure fields \cite{tinney2008low, jonathan2013integration,chen2022pressure}. Nevertheless, their reliance on linear assumptions limits their applicability to complex flows with strongly nonlinear dynamics.

Recent advances in \ac{ML}, particularly in deep learning, have provided powerful alternatives for flow field reconstruction from limited measurements \cite{lusch2018deep, duraisamy2021perspectives}. By learning nonlinear mappings between temporally dense probe signals and the latent dynamics of fluid flows, neural networks are well-suited to capture complex spatio-temporal relationships that are difficult to model using linear techniques. A wide range of architectures have been explored for this purpose, including convolutional neural networks for spatial feature extraction \cite{mo2025reconstructing}, graph neural networks for learning on irregular sensor layouts \cite{danciu2024flow}, and physics-informed neural networks that incorporate governing equations as soft constraints \cite{chaurasia2024reconstruction,morenosoto2024complete}. When combined with reduced-order representations such as those extracted by \ac{POD} or autoencoders, deep learning models can efficiently predict the temporal evolution of dominant flow structures from sparse measurements, offering improved accuracy and robustness over purely linear approaches \cite{eivazi2020deep}.

A key practical observation motivates the present work: in most experimental setups, point probes acquire data at sampling frequencies orders of magnitude higher than standard snapshot \ac{PIV} (typically operating at a few tens of Hz). Consequently, a large amount of probe data is recorded in the time intervals between consecutive \ac{PIV} frames. Such data are not synchronized with velocity fields, and thus lack labels to be used for the supervised training. These probe-only measurements are typically discarded during training (apart from short sequence adjacent to the snapshots, used for time-delay embedding). This results in a severe underutilization of available information.

In this study, we aim to address these limitations by making more effective use of the information provided by high-repetition-rate probe measurements. Two complementary strategies are introduced. First, an advection-based model propagating in time the available velocity fields is employed to enrich the training dataset. By assuming that small-scale flow motions are passively advected by large-scale motions, temporal derivatives of the velocity field can be estimated from individual \ac{PIV} snapshots and used to extrapolate the flow field in time, thereby generating additional training samples near the original snapshots. While limited to advection-dominated cases, this approach has been shown to be successful in the past for pressure estimation \cite{van2019pressure} and time-resolved sequence regularization \cite{chen2025advection}.

Second, a semi-supervised learning framework is developed to explicitly exploit the large amount of unlabelled probe-only data acquired between \ac{PIV} snapshots. The framework trains two neural networks to predict the temporal coefficients of \ac{POD} modes and their temporal derivatives from multi-time-delay embeddings of probe signals. While labelled samples provide direct supervision, unlabelled samples introduce temporal consistency constraints that regularize the learning process and expand the range of flow scenarios represented in the training data. To further improve accuracy, a synthesis step based on least-squares minimization is introduced to reconcile the predictions of the two networks and enforce consistency between the reconstructed temporal coefficients and their temporal derivatives.

The proposed method is validated on both synthetic and experimental datasets, including a turbulent channel flow and an airfoil wake measured by time-resolved \ac{PIV}. The results demonstrate that incorporating unlabelled probe data significantly improves the accuracy and temporal smoothness of velocity reconstruction, leading to more reliable pressure estimation via the Navier–Stokes equations. Importantly, these improvements are achieved without increasing experimental cost, highlighting the potential of semi-supervised learning for data-efficient, probe-based flow reconstruction.

The remainder of this paper is organized as follows. Sec. \ref{sec: methods} introduces the data-driven framework for flow estimation from probe measurements, including a brief review of \ac{EPOD}-based reconstruction and its deep-learning-based extensions. The strategy for enriching the training dataset via velocity-field propagation is then presented, followed by the formulation of the semi-supervised learning approach that exploits overabundant unlabelled probe data. Sec. \ref{sec: strategy} describes the training strategy, including the training procedures and the post-processing of the deep learning model outputs. The proposed method is first validated on a synthetic turbulent channel flow case in Sec. \ref{sec: synthetic}, and subsequently applied to an experimental airfoil wake dataset in Sec. \ref{sec: experimental}. Finally, the main findings and conclusions are summarized in Sec. \ref{sec: conclusion}.

\section{Methods}
\label{sec: methods}

\subsection{Data-driven flow estimation from probe data}

The data-driven flow estimation method presented in this paper leverages high-repetition-rate probe data to enhance the temporal resolution of \ac{PIV} fields. This approach involves two stages: training and testing. During the training phase, velocity fields (e.g., obtained via \ac{PIV}) and probe signals are simultaneously captured, and a model is trained to establish a mapping from probe to field data. In the testing phase, the time-resolved flow fields are reconstructed from the probe data using the trained model and tested.

As done in previous studies \cite{chen2022pressure,chen2025advection}, under the hypothesis of advection-dominated flow, the probes are strategically placed downstream of the \ac{PIV} \ac{FOV} to minimize flow interference. Information from upstream regions is captured using probe data with time-delay embedding \cite{ewing1999examination,tinney2008low}, in which segments of probe signals recorded after each snapshot emulate virtual probes. These virtual probes effectively cover the upstream-to-downstream extent of the \ac{FOV} in advection-dominated flows by exploiting the space–time relationship. The multiple virtual-probe channels further enable the model to predict different flow states.


\subsubsection{EPOD as a linear baseline}
While the proposed framework relies on neural-network estimators, we briefly introduce here the \ac{EPOD} \citep{boree2003extended} framework as a baseline linear tool to establish the correlation between different quantities. Consider a dataset of $n_t$ snapshots of the velocity field $\mathbf{U}(\mathbf{x},t)$. The velocity field is defined on $n_p$ grid points with $n_c$ components for each grid point. A snapshot matrix $\mathbf{U}_{train}(\mathbf{x},t)\in\mathbb{R}^{n_t\times n_{c}n_{p}}$ is built by arranging each snapshot as a row vector. Following the snapshot method from Ref.~\cite{sirovich1987turbulence}, the \ac{POD} is obtained by decomposing $\mathbf{U}_{train}(\mathbf{x},t)$ with an economy-size \ac{SVD}, i.e.
\begin{equation}
    \mathbf{U}_{train}(\mathbf{x},t) - \mathbf{U}_{mean}(\mathbf{x}) = \mathbf{\Psi}(t)\mathbf{\Sigma}\mathbf{\Phi}^T(\mathbf{x})\,,
    \label{eq: POD_U}
\end{equation}
where $\mathbf{U}_{mean}(\mathbf{x})$ represents the matrix of ensemble mean velocity, calculated by averaging $\mathbf{U}_{train}(\mathbf{x},t)$ along its first dimension to produce a $1 \times n_c n_p$ vector. 
Assuming $n_t<n_cn_p$ for simplicity, the decomposition of Eq. \ref{eq: POD_U} generates two square unitary matrices: $\mathbf{\Psi}(t)$, of size $n_t\times n_t$, containing the temporal modes; $\mathbf{\Phi}^T(\mathbf{x})$ containing the spatial modes, of size $n_t \times n_cn_p$. The diagonal matrix $\mathbf{\Sigma}$ contains the singular values $\sigma_i$ sorted by their magnitude.

For any velocity field sample not included in the training dataset $\mathbf{U}_{ext}(\mathbf{x},t)$, the relevant temporal coefficients $\mathbf{\Psi}_{ext}(t)$ can be obtained by projecting it onto the \ac{POD} spatial basis,
\begin{equation}
    \mathbf{\Psi}_{ext}(t)=\big(\mathbf{U}_{ext}(\mathbf{x},t) - \mathbf{U}_{mean}(\mathbf{x})\big)\mathbf{\Phi}(\mathbf{x})\mathbf{\Sigma}^{-1}\,.
    \label{eq: projection}
\end{equation}
Conversely, if an estimation of the temporal coefficients is available, it can be used to reconstruct the velocity fields,
\begin{equation}
    \mathbf{U}_{ext}(\mathbf{x},t) = \mathbf{\Psi}_{ext}(t)\mathbf{\Sigma}\mathbf{\Phi}^T(\mathbf{x}) + \mathbf{U}_{mean}(\mathbf{x}) + \mathbf{r}(\mathbf{x},t)\,.
    \label{eq: decoupled}
\end{equation}
Here, $\mathbf{r}(\mathbf{x},t)$ denotes the part of $\mathbf{U}_{ext}(\mathbf{x},t)$ that is perpendicular to the linear subspace spanned by column vectors of $\mathbf{\Phi}(\mathbf{x})$, viz., $\mathbf{r}(\mathbf{x},t)\perp\mathcal{L}(\mathbf{\Phi}(\mathbf{x}))$, which is referred as decorrelated part in \cite{boree2003extended}. If $\mathbf{U}_{ext}(\mathbf{x},t)$ are well correlated with the training dataset, they should fall into $\mathcal{L}(\mathbf{\Phi}(\mathbf{x}))$, meaning that $\mathbf{r}(\mathbf{x},t)$ would be negligible. This gives the opportunity to estimate the velocity field $\mathbf{U}_{ext}(\mathbf{x},t)$ in the low-dimensional state determined by the encoded temporal coefficients $\mathbf{\Psi}_{ext}(t)$.

An effective implementation of the estimation can be developed from the \ac{EPOD}, by building the linear correlation between the temporal modes of the velocity fields and of probe signals \cite{boree2003extended,tinney2008low,discetti2018estimation}, thus providing a pathway to estimate the time coefficients $\mathbf{\Psi}_{ext}(t)$. As mentioned above, the number of probes can be artificially increased by a multi-time-delay embedding, i.e., considering for each snapshot a time segment of the probe data for time instants immediately past the \ac{PIV} snapshot. In this arrangement, with probes located downstream and the flow dominated by advection, the sensed data would inform of events happening upstream the sensor location at the instant in which the \ac{PIV} snapshot was captured. A multi-time-delay embedding of $q$ time samples results in a probe snapshot matrix $\mathbf{s}_{train}(t)$ with $n_t$ rows and $n_{tt}=(n_s\times n_{c_{pr}})\times q$ columns (being $n_s$ the number of probes and $n_{c_{pr}}$ the number of flow quantity components measured by the probes). This matrix undergoes the same decomposition of the velocity snapshot matrix, with $\mathbf{s}_{mean}$ being the ensemble average recorded by each probe (both real and embedded), resulting in
\begin{equation}
    \mathbf{s}_{train}(t) - \mathbf{s}_{mean} = \mathbf{\Psi}_{s}\mathbf{\Sigma}_{s}\mathbf{\Phi}_{s}^T\,,
    \label{eq: POD_P}
\end{equation}

\noindent with the subscript $s$ indicating probe data. The correlation matrix of temporal modes is thus built as $\mathbf{\Xi}=\mathbf{\Psi}_{s}^T\mathbf{\Psi}$. The matrix $\mathbf{\Xi}$ establishes the correlation between the velocity field and probe temporal modes. The estimated temporal modes of the velocity field $\mathbf{\Psi}_{test}(t)$, of size $n_{test}\times n_t$ (with the subscript ${test}$ indicating testing snapshots), can be a function of probe data used for testing $\mathbf{s}_{test}(t)$,
\begin{equation}
    \mathbf{\Psi}_{test}(t) = (\mathbf{s}_{test}(t)-\mathbf{s}_{mean})\mathbf{\Phi}_{s}\mathbf{\Sigma}_{s}^{-1}\mathbf{\Xi}\,.
    \label{eq: EPOD}
\end{equation}

In some applications \cite{chen2022pressure}, low-pass filters are applied to $\mathbf{\Psi}_{test}(t)$ to suppress the fluctuation of the estimated temporal modes, and improve the temporal derivatives (which is relevant, for instance, for pressure field estimation). After that, the velocity field can be reconstructed using Eq. \ref{eq: decoupled},
\begin{equation}
    \mathbf{U}_{test}(\mathbf{x},t) = \mathbf{\Psi}_{test}(t)\mathbf{\Sigma}\mathbf{\Phi}^T(\mathbf{x}) + \mathbf{U}_{mean}(\mathbf{x})\,.
    \label{eq: recon}
\end{equation}

The reconstruction sketched so far is a well-assessed linear method to enhance the temporal resolution of the \ac{PIV} fields with high-repetition-rate probes, which has been proven to be effective for the flows with compact \ac{POD} spectrum and dominance of advection. Based on this method, Chen et al. \cite{chen2022pressure} showed that the pressure fields can be computed from the time-resolved velocity fields via the Navier-Stokes equations without introducing further assumptions on the temporal derivatives. 

\subsubsection{The nonlinear framework for flow estimation}
In this work, we note a non-linear mapping between probe signal $\mathbf{s}(t)$ and \ac{POD} coefficient $\mathbf{\Psi}(t)$ can be established using supervised \ac{ML} (or briefly, SML) strategies in order to capture the nonlinear part of the mapping between the datasets. The model is represented as
\begin{equation}
    \hat{\mathbf{\Psi}}(t) = \mathbf{f}\left(\mathbf{s}(t)\right)\,,
\end{equation}
where $\hat{\mathbf{\Psi}}(t)$ is the $1\times n_t$ vector containing the predicted temporal coefficients at time $t$ and $\mathbf{f}$ is an optimal non-linear mapping which best approximates the prediction to the measured time coefficients $\mathbf{\Psi}(t)$ according to,
\begin{equation}
    \mathbf{f}^* = \mathop{\arg\min}_{\mathbf{f}}\left\|\mathbf{\Sigma}\bigg(\mathbf{\Psi}(t_j)-\mathbf{f}(\mathbf{s}_{train}(t_j))\bigg)^T\right\|, \quad\forall t_j, \text{if } \mathbf{\Psi}(t_j) \text{ is known}\,,
    \label{eq: f_argmin_supervised}
\end{equation}
with $\mathbf{\Psi}(t_j)$ being the vector of temporal coefficients of any sampled velocity field according to Eq. \ref{eq: POD_U}. Notice that Eq. \ref{eq: f_argmin_supervised}  includes the diagonal matrix $\Sigma$ from the POD in Eq. \ref{eq: POD_U}, which works as a weighting factor for the error between $\mathbf{\Psi}$ and $\mathbf{f}(\mathbf{s}_{train})$, and the weights progressively decrease from lower- to higher-order \ac{POD} modes. The operator $\|\cdot\|$ represents an error metric to be specified (being in this work and in most cases based on an $L^1$ norm). In practice, the prediction of $\hat{\mathbf{\Psi}}(t)$ is often truncated to the first $M$ modes, since high-order modes are difficult to predict. This yields a compact representation $\hat{\mathbf{\Psi}}_M(t)\in\mathbb{R}^{1\times M}$. From this point onward, the symbol $\mathbf{\Psi}$ will be used to refer exclusively to the truncated version $\mathbf{\Psi}_M$ in \ac{ML}, unless stated otherwise. 

The use of deep learning for the estimation provides several advantages: it can approximate highly nonlinear mappings between probe measurements and modal coefficients, handle high-dimensional input data efficiently, and remain robust to noise and measurement uncertainty. These properties allow the model to infer complex flow dynamics that are difficult to capture using linear methods alone.

However, the training defined in Eq. \ref{eq: f_argmin_supervised} requires a sufficiently larger number of \ac{PIV} snapshots. This is necessary not only to increase the dimensionality of the linear subspace spanned by $\mathbf{\Phi}$ and thereby reduce the residual $\mathbf{r}(\mathbf{x},t)$, but also to ensure that the learning process captures a wider range of possible flow states. Acquiring additional \ac{PIV} snapshots, however, is associated with substantial cost, including increased storage requirements for \ac{PIV} images and significant computational expense for velocity-field estimation. On the other hand, the cost of collecting additional probe data is significantly smaller in terms of storage and processing. Moreover, between consecutive \ac{PIV} snapshots, a large number of unlabelled probe measurements are naturally available. Exploiting these unlabelled probe snapshots helps fill the unobserved temporal gaps between known flow realizations, thereby improving data-driven flow estimation.

\subsection{Expanded training dataset using velocity fields propagation}
\label{sect: enriched}

In order to enrich the training dataset, an advection-based model is adopted that assumes small-scale flow motions are passively advected by the large-scale flow. Under this assumption, the temporal derivative of the velocity field can be estimated directly from single \ac{PIV} snapshots. These estimated derivatives are then used to extrapolate the velocity field beyond the original snapshots, thereby augmenting the training dataset. In detail, the model is based on the assumption that the material derivative of the small-scale velocity fluctuations is negligible, i.e. 
\begin{equation}
    \frac{D\mathbf{U}'}{D t} = \frac{\partial \mathbf{U}'}{\partial t} +(\mathbf{U}_c\cdot\nabla)\mathbf{U}' \approx 0\,,
\end{equation}
where the $\mathbf{U}_c$ denotes the local convective velocity, and $\mathbf{U}'$ represents $\mathbf{U}-\mathbf{U}_c$. The convective velocity can be estimated using various approaches, ranging from simple temporal (or spatial) -averaged velocity profiles to more advanced techniques such as spatio-temporal correlation analysis \cite{del2009estimation}. In this study, $\mathbf{U}_c$ is determined using a spatial low-pass filter combined with, if applicable, a near-wall correction, as in \cite{chen2025advection}.
Assuming that the timescale of large-scale motions is significantly larger than that of small-scale motions, the temporal derivative of the velocity field can be approximated as,
\begin{equation}
    \frac{\partial\mathbf{U}}{\partial t}\approx\frac{\partial\mathbf{U}'}{\partial t}\approx-(\mathbf{U}_c\cdot\nabla)\mathbf{U}'.
    \label{eqn:TH}
\end{equation}
This method has been previously applied to pressure estimation from snapshot \ac{PIV} data to supplement the temporal derivative of the velocity field in the Navier–Stokes equations \cite{A32019pressure}. Here, Eq. \ref{eqn:TH} is used to propagate the velocity field forward or backward to the instant where only probe data are available. The step number for the propagation while maintaining accuracy depends on several factors, such as the repetition rate of the probes, the shear strength of the flow, and the motion in directions other than the streamwise direction. In practice, propagating the velocity field until $1/20$ convective time will bring positive effect in data-driven estimation than introduce the error from distorted propagation or information vacuum in the upstream/downstream. A 4-th order Runge–Kutta scheme is used for this purpose. The temporal coefficients  $\mathbf{\Psi}(t)$ of the propagated snapshots are obtained by means of projection according to Eq. \ref{eq: projection}. This procedure increases the number of training samples for supervised learning. In the rest of this paper, training with the expanded dataset from propagation will be called supervised \ac{ML} on expanded dataset – or, for short, supervised ML on Ex. (abbreviated SML on Ex.).

\subsection{Semi-supervised deep learning}
\label{sect: semi-supervised}

The semi-supervised \ac{ML} (or briefly, SSML) strategy is based on the hypothesis that the velocity field $\mathbf{U}(\mathbf{x},t)$ in Eq. \ref{eq: decoupled} is temporally differentiable throughout the entire domain, implying the absence of phenomena such as moving objects, shock waves, or expansion regions. Notably, since $\mathbf{\Sigma}\mathbf{\Phi}^T(\mathbf{x})$ is composed of orthogonal row vectors, there exists a unique set of coefficients $c_1(t_0), c_2(t_0), ..., c_{n_t}(t_0)$ at any time instant $t_0$ satisfying,
\begin{equation}
    \mathbf{U}_t(\mathbf{x},t_0) = \mathop{\lim}_{\Delta t\rightarrow0}\frac{\mathbf{U}(\mathbf{x},t_0+\Delta t)-\mathbf{U}(\mathbf{x},t_0)}{\Delta t}
    = [c_1(t_0), c_2(t_0), ..., c_{n_t}(t_0)]\mathbf{\Sigma}\mathbf{\Phi}^T(\mathbf{x})
\end{equation}
with 
\begin{equation}
     c_i(t_0)=\mathop{\lim}_{\Delta t\rightarrow0}\frac{\psi_i(t_0+\Delta t)-\psi_i(t_0)} {\Delta t},\qquad\forall i\in[1,n_t]\,,
\end{equation}
where $\mathbf{U}_t(\mathbf{x},t)$ denotes the temporal derivative of the velocity field $\mathbf{U}(\mathbf{x},t)$, and $\psi_i(t_0)$ denotes the $i^{th}$ component of the temporal coefficient vector $\mathbf{\Psi}(t_0)$. Under these definitions, the temporal derivative of $\mathbf{\Psi}(t)$ in Eq. \ref{eq: decoupled} exists \footnote{In this paper, \ac{POD} spatial modes are used for encoding and decoding the velocity field. More advanced approaches may employ neural-network-based autoencoders to map the velocity field into a latent space and reconstruct it thereafter. A generic encoder can be expressed as $\mathbf{z}(\mathbf{U})=\phi_L(\mathbf{W}_L\phi_{L-1}(\mathbf{W}_L...\phi_1(\mathbf{W}_1\mathbf{U}+\mathbf{b}_1)+\mathbf{b}_{L-1})+\mathbf{b}_L)$, with $\mathbf{W}_i$, where $\mathbf{W}_i$ and $\mathbf{b}_i$ denote the weights and biases of each layer, and $\phi_i$ are the activation functions. If the activation functions are differentiable everywhere, the encoded latent variables $\mathbf{z}$ inherit the temporal differentiability of $\mathbf{U}$. Conversely, non-differentiable activation functions may compromise the differentiability of $\mathbf{z}$ at specific points. Further discussion of activation functions can be found in \cite{apicella2021survey}. When the latent space is treated probabilistically, the analysis becomes more involved; however, since velocity fields are typically discretized in time, such approaches often remain practical.}. We denote this temporal derivative by $\mathbf{\Psi}_t(t)$ throughout this paper. It then follows that $\mathbf{U}_t(\mathbf{x},t)$ can be reconstructed as
\begin{equation}
    \mathbf{U}_t(\mathbf{x},t) = \mathbf{\Psi}_t(t)\mathbf{\Sigma}\mathbf{\Phi}^T(\mathbf{x})\,.
    \label{eq: recon_ut}
\end{equation}

\begin{figure}[htbp]
\centering
\includegraphics[width=0.5\textwidth, trim=0 0 0 0, clip]{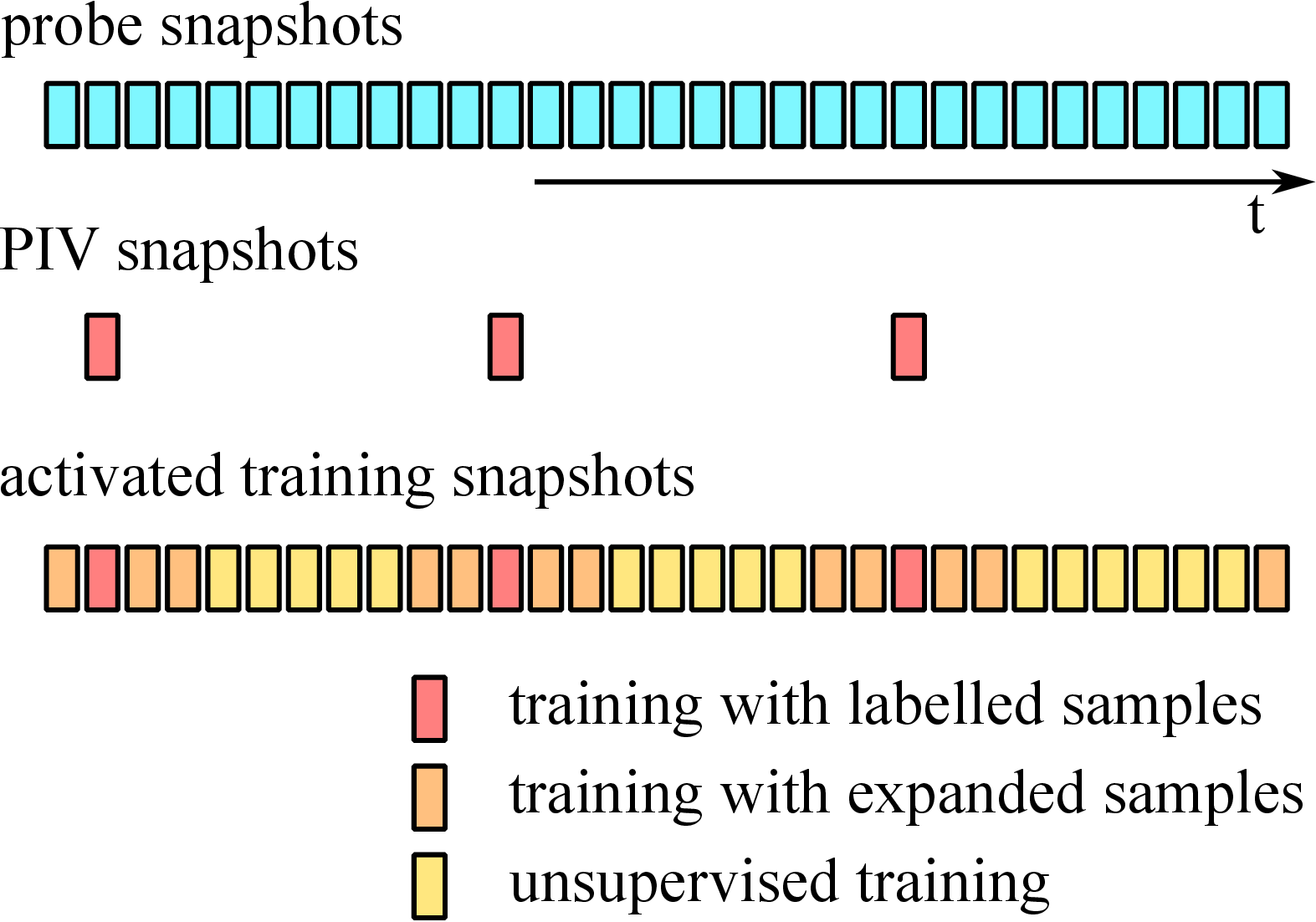}
\caption{Sketch of the different use of data during training, including dataset expansion with advection hypothesis, and unsupervised training for unlabelled snapshots.}\label{fig: data_usage}
\end{figure}

If the model $\mathbf{f}$ in Eq. \ref{eq: f_argmin_supervised} is adequately trained, then for any time instant $t_j$, the prediction error $\left\|\mathbf{\Psi}(t_j)-\mathbf{f}(\mathbf{s}(t_j))\right\|$ should be small. This accuracy is expected to extend to neighbouring time instants $t_{j+k}$, with $k = 0, \pm1, \pm2, \ldots$. Motivated by this temporal consistency, an additional model $\mathbf{g}$ is introduced to predict the temporal derivative $\mathbf{\Psi}_t$, i.e.
\begin{equation}
    \hat{\mathbf{\Psi}}_t(t) = \mathbf{g}\left(\mathbf{s}(t)\right)\,,
\end{equation}
where the model $\mathbf{g}$ is optimized by the finite difference of the prediction on $\mathbf{f}$,
\begin{equation}
    \mathbf{g}^* = \mathop{\arg\min}_{\mathbf{g}}\left\|\mathbf{\Sigma}\left(\frac{\mathbf{f}\left(\mathbf{s}(t_{j+1})\right)-\mathbf{f}\left(\mathbf{s}(t_{j-1})\right)}{t_{j+1}-t_{j-1}}-\mathbf{g}(\mathbf{s}(t_j))\right)^T\right\|, \quad \forall t_j\,.
    \label{eq: g_argmin}
\end{equation}
This approach applies to any probe snapshot at time $t_j$, provided that the flow conditions are similar to those represented in the training of Eq. \ref{eq: f_argmin_supervised}, regardless of whether the probe signal $\mathbf{s}(t_j)$ is labelled with the corresponding temporal modes $\mathbf{\Psi}(t_j)$ projected from \ac{PIV} snapshots. Moreover, the model $\mathbf{f}$ can be further refined using the predictions of $\mathbf{g}$, and the semi-supervised \ac{ML} is fulfilled by both the supervised and unsupervised processes,
\begin{equation}
    \mathbf{f}^* = \left\{
    \begin{array}{ll}
        \underset{\mathbf{f}}{\mathop{\arg\min}} \displaystyle 
        \Bigg(C_1\Bigg\| 
            \mathbf{\Sigma}\Big(\mathbf{\Psi}(t_j) - \mathbf{f}(\mathbf{s}(t_j))\Big)^T \Bigg\|\\
            \quad\quad\quad\,\, + \displaystyle C_2\Bigg\|\mathbf{\Sigma}\Bigg(
                \frac{\mathbf{f}\left(\mathbf{s}(t_{j+1})\right) - \mathbf{f}\left(\mathbf{s}(t_{j-1})\right)}{t_{j+1}-t_{j-1}} 
                - \mathbf{g}(\mathbf{s}(t_j))
            \Bigg)^T
        \Bigg\|\Bigg), & \forall t_j, \text{if } \mathbf{\Psi}(t_j) \text{ is known}, \\[3ex]
        \underset{\mathbf{f}}{\mathop{\arg\min}} \displaystyle 
        \Bigg(C_2\Bigg\| 
            \mathbf{\Sigma}\Bigg(
                \frac{\mathbf{f}\left(\mathbf{s}(t_{j+1})\right) - \mathbf{f}\left(\mathbf{s}(t_{j-1})\right)}{t_{j+1}-t_{j-1}} 
                - \mathbf{g}(\mathbf{s}(t_j))
            \Bigg)^T
        \Bigg\|\Bigg), & \forall t_j, \text{if } \mathbf{\Psi}(t_j) \text{ is unknown}.
    \end{array}
    \right.
    \label{eq: f_argmin}
\end{equation}
Here, $C_1$ and $C_2$ are hyper-parameters to be tuned. The semi-supervised deep learning framework leverages not only probe samples labelled by \ac{PIV} snapshots but also unlabelled probe samples, enabling more efficient usage of high-repetition-rate probe data, as plotted in Fig. \ref{fig: data_usage}. This approach improves the model performance and progressively expands its range of applicability.

\section{Training strategy}
\label{sec: strategy}

\subsection{Training procedures}

In this section, we provide a formal definition of the dataset, detail the training procedure, and how the outputs of both neural networks are used.

Several sets of samples are defined prior to training:
\begin{enumerate}
    \item \textbf{Supervised-training set:} Probe samples with corresponding \ac{PIV} snapshots at the same time instants, which are used for supervised training in Eq. \ref{eq: f_argmin_supervised}. The corresponding time indices are denoted as $T_S$.
    \item \textbf{Expanded-training set:} Probe snapshots without directly corresponding \ac{PIV} snapshots, where the associated $\mathbf{\Psi}(t_j)$ can be obtained by propagation from nearby \ac{PIV} snapshots using the method described in Sect. \ref{sect: enriched}. The corresponding time indices are denoted as $T_P$.
    \item \textbf{Unsupervised-training set:} Probe snapshots without corresponding \ac{PIV} snapshots, used exclusively for unsupervised training. Their time indices are denoted as $T_U$.
    \item \textbf{Validation set:} A set of synchronized probe and field snapshots reserved for the validation of the mapping operators. The corresponding time indices are denoted as $T_V$.
    \item \textbf{Testing set:} Probe samples used for testing, with the corresponding time indices denoted as $T_T$.
\end{enumerate}

The training process consists of three stages. In the first stage, the model $\mathbf{f}(\mathbf{s}(t))$ is trained using only the supervised-training and expanded-training sets, with the loss function defined as
\begin{equation}
\begin{aligned}
    &L_1 = K_1 + C_{11}P_1, \qquad\forall t_j\in T_S\cup T_P\,, \\
    &\text{where } K_1=\left\|\mathbf{\Sigma}\Big(\mathbf{\Psi}(t_j) - \mathbf{f}(\mathbf{s}(t_j))\Big)^T\right\|_1\\
    &\qquad\,\,\,\,\, P_1 = \,\, \left\|\mathbf{f}(\mathbf{s}(t_{j-1}))-2\mathbf{f}(\mathbf{s}(t_j))+\mathbf{f}(\mathbf{s}(t_{j+1}))\right\|_2\,.
\end{aligned}
\label{eq: stage1}
\end{equation}
Here, the \ac{MAE} $\|\cdot\|_1$ is used to reduce sensitivity to outliers and prevent the predicted $\mathbf{\Psi}$ from being biased toward zero. Meanwhile, the penalty term $P_1$, based on the \ac{MSE} $\|\cdot\|_2$, is applied to suppress fluctuations of $\mathbf{f}(\mathbf{s}(t))$ over time, improving the stability of $\mathbf{f}(\mathbf{s}(t))$ predictions for the subsequent training stage, which uses finite differences of $\mathbf{f}(\mathbf{s}(t))$ over $t$. The coefficient $C_{11}$ in this equation is a hyperparameter to be tuned, similarly to coefficients of the form $C_{nn}$ in the rest of this section, whose value will be listed in Sec. \ref{sec: synthetic} and \ref{sec: experimental}.

In the second stage, the model $\mathbf{g}(\mathbf{s}(t))$ is trained on all the available probe data (either labelled or unlabelled)   to predict the temporal derivatives of the coefficients, $\mathbf{\Psi}_t$, leveraging the model prediction obtained from the first stage. The corresponding loss function is defined as follows:
\begin{equation}
\begin{aligned}
    &L_2 = K_2 + C_{21}P_2, \qquad\forall t_j\in T_S\cup T_P\cup T_U, \\
    &\text{where } K_2=\left\|\mathbf{\Sigma}\Bigg(
                \frac{\mathbf{f}\left(\mathbf{s}(t_{j+1})\right) - \mathbf{f}\left(\mathbf{s}(t_{j-1})\right)}{t_{j+1}-t_{j-1}} 
                - \mathbf{g}(\mathbf{s}(t_j))
            \Bigg)^T\right\|_1\\
    &\qquad\,\,\,\,\, P_2 = \,\, \left\|\mathbf{g}(\mathbf{s}(t_{j-1}))-2\mathbf{g}(\mathbf{s}(t_j))+\mathbf{g}(\mathbf{s}(t_{j+1}))\right\|_2
\end{aligned}
\label{eq: stage2}
\end{equation}
Here, the \ac{MAE} is employed for the same reasons as in the first stage. The penalty term $P_2$ is introduced to suppress temporal fluctuations of $\mathbf{g}(\mathbf{s}(t))$, thereby promoting smoother and more stable predictions. A two-point finite central differences with the minimum time stencil is applied here, optionally different schemes can be used, such as Savitzky-Golay scheme \cite{savitzky1964smoothing} to reduce the random error.

In the third stage, training is refined using both labelled and unlabelled data with the corresponding loss function. This stage is used to fine-tune the models, further improving their accuracy and dynamic performance.

\begin{equation}
\begin{aligned}
    &L_3 \,= \left\{
    \begin{aligned}
        K_1 + &C_{31}\tilde{K}_2,\qquad \forall t_j\in T_S\cup T_P, \\               
         &C_{31}\tilde{K}_2,\qquad \forall t_j\in T_U.
    \end{aligned}
    \right.\\
    &L_{3t} = \,\,\,K_2 + C_{32}P_2,  \qquad    \forall t_j\in T_S\cup T_P\cup T_U\\
    &\text{where } \tilde{K}_2=\left\|\mathbf{\Sigma}\Bigg(
                \frac{\mathbf{f}\left(\mathbf{s}(t_{j+1})\right) - \mathbf{f}\left(\mathbf{s}(t_{j-1})\right)}{t_{j+1}-t_{j-1}} 
                - \mathbf{g}(\mathbf{s}(t_j))
            \Bigg)^T\right\|_2\\
\end{aligned}
\label{eq: stage3}
\end{equation}

The loss function $L_3$ is used to refine the model $\mathbf{f}(\mathbf{s})$, during which the weights of $\mathbf{f}(\mathbf{s})$ are updated while the weights of $\mathbf{g}(\mathbf{s})$ are kept fixed. Conversely, the loss function $L_{3t}$ is used to train $\mathbf{g}(\mathbf{s})$, with the weights of $\mathbf{g}(\mathbf{s})$ updated and those of $\mathbf{f}(\mathbf{s})$ held fixed. The training of $\mathbf{f}(\mathbf{s})$ and $\mathbf{g}(\mathbf{s})$ is performed alternately.

Validation plays a pivotal role in fine-tuning deep learning models and their hyper-parameters, as well as identifying overfitting or underfitting. In this study, validation is also essential for integrating the outputs of the two models, as discussed later in Sect. \ref{sect: synthesis}. The following loss functions are used for validation, both based on the \ac{MSE} to facilitate consistent evaluation,

\begin{equation}
\begin{aligned}
    L_{3,val} &=\displaystyle\Bigg\| 
            \mathbf{\Sigma}\Big(\mathbf{\Psi}(t_j) - \mathbf{f}(\mathbf{s}(t_j))\Big)^T \Bigg\|_2, & \forall t_j\in T_V, \\[3ex]
    L_{3t,val} &= \left\|\mathbf{\Sigma}\left(\frac{\mathbf{f}\left(\mathbf{s}(t_{j+1})\right)-\mathbf{f}\left(\mathbf{s}(t_{j-1})\right)}{t_{j+1}-t_{j-1}}-\mathbf{g}(\mathbf{s}(t_j))\right)^T\right\|_2, & \forall t_j\in T_V
\end{aligned}
\label{eq: loss_validation}
\end{equation}

\subsection{Post-processing of deep learning model outputs}
\label{sect: synthesis}

Applying the model $\mathbf{f}(\mathbf{s}(t))$ and $\mathbf{g}(\mathbf{s}(t))$ to the testing dataset $T_T$ yields two outputs: the predicted temporal coefficients $\hat{\mathbf{\Psi}}\in\mathbb{R}^{n_{test}\times M}$, and their predicted temporal derivative $\hat{\mathbf{\Psi}}_t\in\mathbb{R}^{n_{test}\times M}$. These quantities correspond to the reconstructed velocity fields and their temporal derivative through the basis $\mathbf{\Sigma}\mathbf{\Phi}^T$, as defined in Eqs. \ref{eq: recon} and \ref{eq: recon_ut}.

Since $\hat{\mathbf{\Psi}}$ and $\hat{\mathbf{\Psi}}_t$ are predicted by separate models, the temporal derivative of $\hat{\mathbf{\Psi}}$ in numerical generally does not necessarily coincide with $\hat{\mathbf{\Psi}}_t$, even though the models $\mathbf{f}$ and $\mathbf{g}$ are trained to satisfy the constraints in Eqs. \ref{eq: g_argmin} and \ref{eq: f_argmin}. To reconcile this discrepancy and enforce temporal consistency, a synthesis step based on the \ac{LSM} is employed.

The objective of the post-processing step is to regularise the estimates of the temporal coefficients $\mathbf{\Psi}_{LS}$ and their 
temporal derivatives $\mathbf{\Psi}_{t,LS}$ by minimizing weighted 
residuals, subject to the constraint that the numerical time derivative of $\mathbf{\Psi}$ coincides 
with $\mathbf{\Psi}_t$. The quantity to be minimized is:
\begin{equation}
    S = \sum_{i,j}\left(\alpha_1^2\left(\sigma_j(\psi_{i,j}-\hat{\psi}_{i,j})\right)^2 + \alpha_2^2\left(\sigma_j(\psi_{t,i,j}-\hat{\psi}_{t,i,j})\right)^2\right)
    \label{eq: LSM1}
\end{equation}
where $\hat{\psi}_{i,j}$, $\psi_{i,j}$, $\hat{\psi}_{t,i,j}$, and $\psi_{t,i,j}$ denote the elements in the $i^{th}$ row and $j^{th}$ column of the matrices $\hat{\mathbf{\Psi}}$, $\mathbf{\Psi}$, $\hat{\mathbf{\Psi}}_t$, and $\mathbf{\Psi}_t$, respectively, with the indices $i$ and $j$ corresponding to the time index and mode number. The quantity $\sigma_j$ denotes the $j^{th}$ element of $\operatorname{diag}(\mathbf{\Sigma})$. The coefficients $\alpha_1^2$ and $\alpha_2^2$ are weighting parameters to be determined, chosen such that more reliable estimates exert greater influence on the final result.

Generalized least squares theory prescribes that the coefficients $\alpha_1^2$ and $\alpha_2^2$ should be chosen inversely to the variances of the residuals $\sigma_j(\psi_{i,j}-\hat{\psi}_{i,j})$ and $\sigma_j(\psi_{t,i,j}-\hat{\psi}_{t,i,j})$, respectively \cite{aitken1936iv}. Assuming the improved estimation $\mathbf{\Psi}_{LS}$ and $\mathbf{\Psi}_{t,LS}$ fall close to the ground truth, we approximate the variance of $\sigma_j(\psi_{i,j}-\hat{\psi}_{i,j})$ using the validation loss $L_{3,\mathrm{val}}$ defined in Eq. \ref{eq: loss_validation}. Estimating the variance of $\sigma_j(\psi_{t,i,j}-\hat{\psi}_{t,i,j})$ is more involved, since $\psi_{t,i,j}$ cannot be obtained directly. We instead approximate it through the upper bound of $L_{3t,val}$ obtained via the triangle inequality,
\begin{equation}
\begin{aligned}
    L_{3t,val}\leq& \left\|\mathbf{\Sigma}\left(\mathbf{\Psi}_t(t_j)-\frac{\mathbf{f}\left(\mathbf{s}(t_{j+1})\right)-\mathbf{f}\left(\mathbf{s}(t_{j-1})\right)}{t_{j+1}-t_{j-1}}\right)^T\right\|_2 + \Bigg\|\mathbf{\Sigma}\bigg(\mathbf{\Psi}_t(t_j)-\mathbf{g}(\mathbf{s}(t_j))\bigg)^T\Bigg\|_2, & \forall t_j\in T_V
\end{aligned}
\label{eq: L3tv_split}
\end{equation}
In our experiments, the first term on the right hand side is significantly larger than the second term due to the numerical differentiation error. We therefore introduce a scaling factor $C_\alpha=8$ determined empirically to approximate $\|\mathbf{\Sigma}(\mathbf{\Psi}_t(t_j)-\mathbf{g}(\mathbf{s}(t_j)))^T\|_2$ as $L_{3t,val} = C_\alpha\|\mathbf{\Sigma}(\mathbf{\Psi}_t(t_j)-\mathbf{g}(\mathbf{s}(t_j)))^T\|_2$, and the $\alpha_1^2$ and $\alpha_2^2$ are then given by,
\begin{equation}
    \begin{aligned}
        \alpha_1^2 &= 1/L_{3,val}\\
        \alpha_2^2 &= C_\alpha/L_{3t,val}
    \end{aligned}
\end{equation}

The matrix $\hat{\mathbf{\Psi}}$ is reshaped to a vector $ \operatorname{vec}(\hat{\mathbf{\Psi}}) \in \mathbb{R}^{M\,n_{test} \times 1}$ by concatenating its column vectors into a single column vector (in column-major order). The same reshaping is applied to $\mathbf{\Psi}$, $\hat{\mathbf{\Psi}}_t$, and $\mathbf{\Psi}_t$, resulting in four vectors, each of size $M\,n_{test} \times 1$. The temporal derivative constraint between $\operatorname{vec}(\mathbf{\Psi}_t)$ and $\operatorname{vec}(\mathbf{\Psi})$ can then be expressed as,
\begin{equation}
    \operatorname{vec}(\mathbf{\Psi}_t) = \frac{\mathbf{A}}{2\Delta t}\operatorname{vec}(\mathbf{\Psi})
    \label{eq: LSM_constrain}
\end{equation}
where $\Delta t$ denotes the time interval of testing data, the matrix $\mathbf{A}\in\mathbb{R}^{M\,n_{test}\times M\,n_{test}}$ is a finite-difference operator composed of $0$ and $\pm1$, encoding the entries involved in the finite-difference relation. In addition, the matrix $\mathbf{C}\in\mathbb{R}^{Mn_{test}\times Mn_{test}}$ incorporates the weighting factors $\sigma_j$ and is defined as a diagonal matrix. Its diagonal entries are arranged as $\underbrace{\sigma_1, \sigma_1, \ldots, \sigma_1}_{n_{test} \text{ times}}, \underbrace{\sigma_2, \sigma_2, \ldots, \sigma_2}_{n_{test} \text{ times}}, \ldots, \underbrace{\sigma_M, \sigma_M, \ldots, \sigma_M}_{n_{test} \text{ times}}$. With these definitions, Eq. \ref{eq: LSM1} can then be rewritten in matrix form as,
\begin{equation}
    \begin{aligned}
        S =& \alpha_1^2\left\|\mathbf{C}\operatorname{vec}(\mathbf{\Psi})-\mathbf{C}\operatorname{vec}(\hat{\mathbf{\Psi}})\right\|^2 + \alpha_2^2\left\|\frac{\mathbf{C}\mathbf{A}}{2\Delta t}\operatorname{vec}(\mathbf{\Psi})-\mathbf{C}\operatorname{vec}(\hat{\mathbf{\Psi}}_t)\right\|^2\\
        =& \alpha_1^2\left(\operatorname{vec}(\mathbf{\Psi})-\operatorname{vec}(\hat{\mathbf{\Psi}})\right)^T\mathbf{C}^T\mathbf{C}\left(\operatorname{vec}(\mathbf{\Psi})-\operatorname{vec}(\hat{\mathbf{\Psi}})\right)\\
        +& \alpha_2^2\left(\frac{\mathbf{A}}{2\Delta t}\operatorname{vec}(\mathbf{\Psi})-\operatorname{vec}(\hat{\mathbf{\Psi}}_t)\right)^T\mathbf{C}^T\mathbf{C}\left(\frac{\mathbf{A}}{2\Delta t}\operatorname{vec}(\mathbf{\Psi})-\operatorname{vec}(\hat{\mathbf{\Psi}}_t)\right)
    \end{aligned}
    \label{eq: LSM2}
\end{equation}
while its gradients by $\operatorname{vec}(\mathbf{\Psi})$ is
\begin{equation}
    \frac{\partial S}{\partial \operatorname{vec}(\mathbf{\Psi})}=2\alpha_1^2\mathbf{C}^T\mathbf{C}\operatorname{vec}(\mathbf{\Psi}) - 2\alpha_1^2\mathbf{C}^T\mathbf{C}\operatorname{vec}(\hat{\mathbf{\Psi}}) + 2\alpha_2^2\frac{\mathbf{A}^T\mathbf{C}^T\mathbf{C}\mathbf{A}}{4\Delta t^2}\operatorname{vec}(\mathbf{\Psi}) - 2\alpha_2^2\frac{\mathbf{A}^T\mathbf{C}^T\mathbf{C}}{2\Delta t}\operatorname{vec}(\hat{\mathbf{\Psi}}_t)
\end{equation}
The minimum of the sum of squares is found by setting the gradients to zero, therefore,
\begin{equation}
    \left(\mathbf{C}^T\mathbf{C}+\frac{\alpha_2^2}{\alpha_1^2}\frac{\mathbf{A}^T\mathbf{C}^T\mathbf{C}\mathbf{A}}{4\Delta t^2}\right)\operatorname{vec}(\mathbf{\Psi}_{LS})=\mathbf{C}^T\mathbf{C}\operatorname{vec}(\hat{\mathbf{\Psi}})+\frac{\alpha_2^2}{\alpha_1^2}\frac{\mathbf{A}^T\mathbf{C}^T\mathbf{C}}{2\Delta t}\operatorname{vec}(\hat{\mathbf{\Psi}}_t)
    \label{eq: LSM_final}
\end{equation}
Solving Eq. \ref{eq: LSM_final} yields $\operatorname{vec}(\mathbf{\Psi}_{LS})$, which minimizes the sum of squared residuals $S$ in Eq. \ref{eq: LSM2}, subject to the constrain in Eq. \ref{eq: LSM_constrain}. The velocity field can then be reconstructed by substituting the reshaped $\mathbf{\Psi}_{LS}$ into Eq. \ref{eq: recon}. Since both $\mathbf{A}$ and $\mathbf{C}$ are highly sparse matrices, 
the computation is generally affordable. In the remainder of this paper, we will denote semi-supervised \ac{ML} before and after \ac{LSM} regularization as SSML‑Reg and SSML+Reg, respectively, for brevity.




\section{Validation}
\label{sec: synthetic}

The proposed deep learning method for probe-based flow prediction is first validated using a synthetic dataset. The assessment is carried out considering both velocity and pressure reconstruction accuracy. The pressure field is computed from the reconstructed velocity field by solving the Navier–Stokes equations for incompressible fluids, in which the temporal derivative of the velocity field is explicitly required,
\begin{equation}
    \frac{\partial\mathbf{u}}{\partial t} + (\mathbf{u}\cdot\nabla)\mathbf{u}=\nu\nabla^2\mathbf{u}-\nabla p
\label{eq: NS}
\end{equation}


The pressure fields are integrated directly from their gradients using an iterative method \cite{tronchin2015loads}. Model training and three-dimensional pressure field computations are conducted on workstations equipped with NVIDIA RTX 3090 GPUs.

\begin{figure}[htbp]
\centering
\includegraphics[width=0.6\textwidth, trim=0 0 0 0, clip]{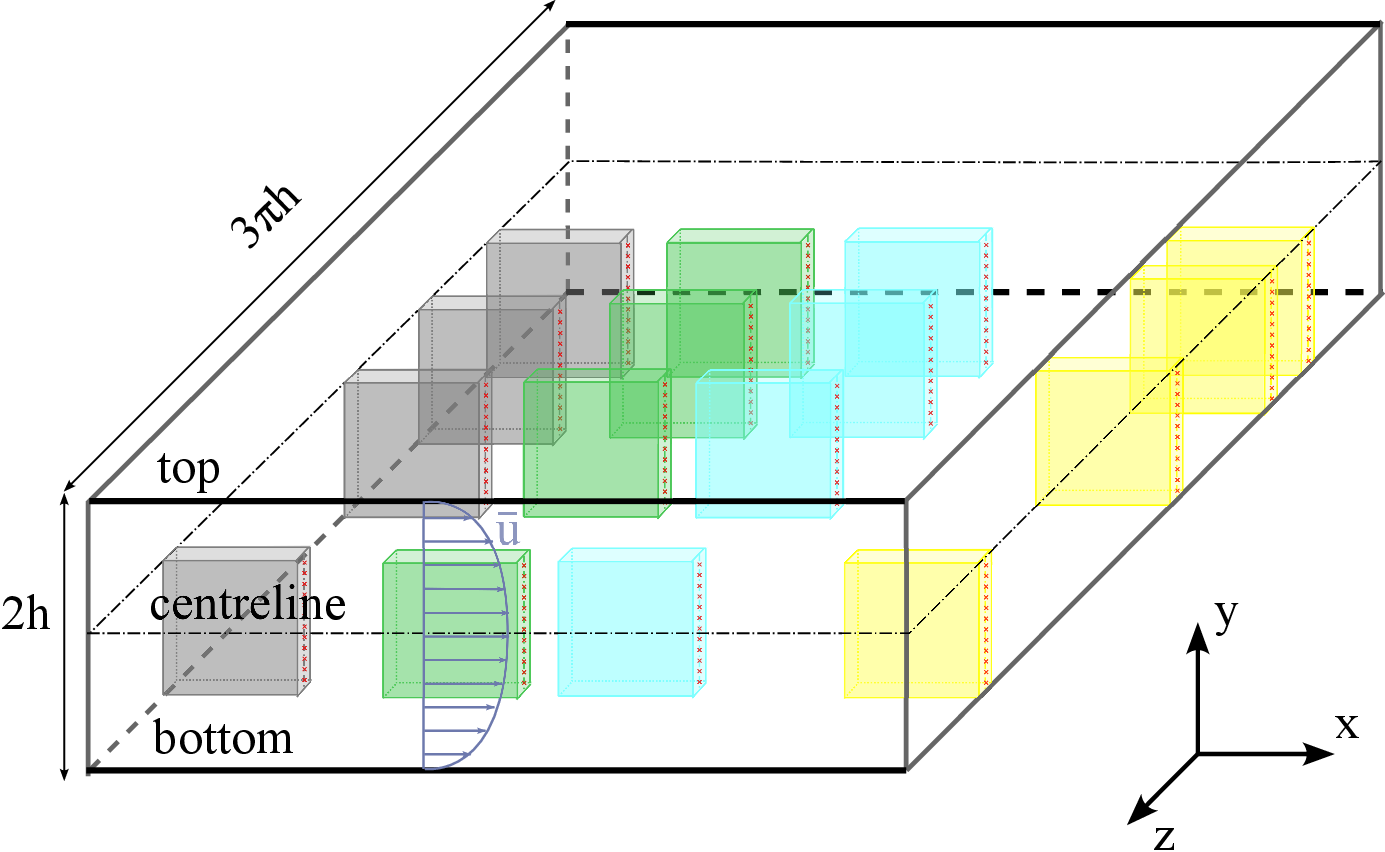}
\caption{Layout of the channel flow dataset. The dyed blocks indicate subdomains, and the red crosses the positions for synthetic probes. The domain is not represented in the full streamwise length of $8\pi h$ for ease of representation.}\label{fig: cketch_channel}
\end{figure}

The synthetic test case for validation is based on a turbulent channel flow simulation. The dataset is resampled from the \ac{DNS} available in the Johns Hopkins Turbulence Databases \cite{li2008public, graham2016web} with a full domain size of $8\pi h \times 2h \times 3\pi h$ in the streamwise, wall-normal, and spanwise directions, respectively, where $h$ denotes the channel half-height. The bulk velocity $U_b$ is $0.9994h$. The friction Reynolds number is $\textit{Re}_\tau \approx 1000$. A three-dimensional subdomain is extracted, at different streamwise locations while preserving the translational invariance of the flow, consisting of $88 \times 88 \times 11$ uniformly spaced grid points and covering a physical region of $h \times h \times h/8$ (Fig. \ref{fig: cketch_channel}). Twelve equidistant probes are placed along the midplane in the spanwise direction on the downstream face of the subdomain. In this study, probe signals are extracted from the high-temporal-resolution streamwise velocity field. For each snapshot, multi-time delay embedding, with the time interval for the probe data to be $\Delta t=0.0065$ unit time as the database, is set up by considering $152$ samples of the probe sequence for each snapshot. This follows the guidelines of Ref.~ \cite{discetti2018estimation}, corresponding to approximately one flow-through time of the subdomain.

The channel-flow dataset consists of $1200$ labelled snapshots, for which both velocity fields and corresponding probe signals are available, used for training and validation, and an additional $500$ frames are reserved for testing. The flow estimation of this dataset applies interpolation mode, the flow fields are estimated between labelled snapshots, whose interval is $24\Delta t$. In addition, the number of randomly chosen unlabelled probe-only samples is $16$ times larger than that of the labelled samples.

\begin{figure}[htbp]
\centering
\includegraphics[width=0.8\textwidth, trim=0 0 30 10, clip]{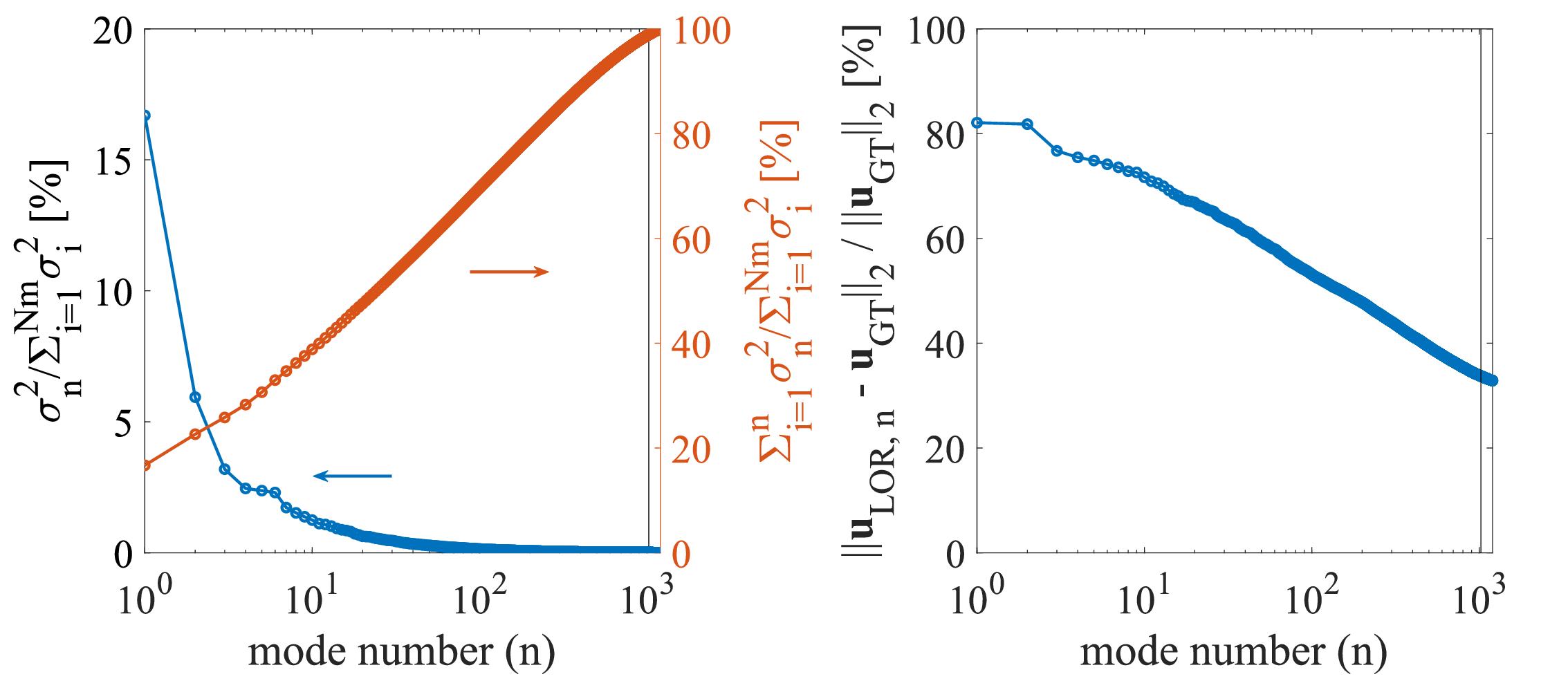}
\caption{Left: the \ac{POD} squared singular values (left), with the descending curve showing the energy contained in each POD mode and ascending curve showing the cumulative energy. Right: Reconstruction error from \ac{LOR} in the testing dataset. The horizontal axis is displayed in logarithmic scale showing the mode numbers.}\label{C_POD}
\end{figure}

\begin{table}[htp]
  \centering
  \caption{The model structure used for the channel dataset}
  \label{tab:channel-arch}
  \begin{tabular}{lcccc}
    \hline
    \textbf{Layer \#} & \textbf{Type}       & \textbf{Output Shape / Units} & \textbf{Activation} & \textbf{Dropout} \\
    \hline
    0 & Input           & $152\times 12$ & ---      & ---    \\
    1 & Dense           & 2048           & tanh     & 10\%   \\
    2 & Reshape         & $(64, 32)$     & ---      & ---    \\
    3 & Dot-Product Self-Attention
                        & $(32, 256)$    & ---      & 10\%   \\
    4 & Flatten         & $2048$         & ---      & ---    \\
    5 & Dense           & 4096           & tanh     & 10\%   \\
    6 & Dense           & 2048           & tanh     & 10\%   \\
    7 & Output Dense    & 1024           & linear   & ---    \\
    \hline
  \end{tabular}
\end{table}

\begin{table}[htp]
\centering
\begin{tabular}{c c c c}
\hline
hyperparameter   & Stage 1          & Stage 2          & Stage 3 \\
\hline
$C_{11}$         & $0$              & --               & --               \\
$C_{21}$         & --               & $0$              & --               \\
$C_{31}$         & --               & --               & $1\times10^{-4}$ \\
$C_{32}$         & --               & --               & $0$              \\
optimizer        & Adam             & Adam             & Adam             \\
learning rate    & $1\times10^{-4}$ & $1\times10^{-4}$ & $5\times10^{-6}$ \\
batch size       & $128$            & $128$            & $128$            \\
number of epochs & $800$            & $200$            & $200$            \\
LP               & $4$              & $4$              & $4$              \\
CP               & $5$              & $5$              & $2$              \\
CU               & --               & $0.2$            & $0.1$            \\
\hline
\end{tabular}
\caption{Hyperparameters used in the channel case. LP: propagation length (number of propagated frames). CP: ratio between propagated and supervised datasets. CU: ratio of (supervised + propagated) to unsupervised data.}
\label{tab:channel-hyper}
\end{table}

The squared singular values of the turbulent channel-flow dataset are shown in Fig. \ref{C_POD}. The left panel presents the \ac{POD} squared singular values and their cumulative sum, expressed as the percentage of the total variance. The right panel illustrates the relative reconstruction error between the \ac{LOR} and the \ac{GT} velocity fields when a representative number of modes is retained for reconstruction. Owing to the high dimensionality and rich dynamics of the turbulent flow, the relative error $\|\mathbf{u}_{LOR, n}-\mathbf{u}_{GT}\|_2$ remains above $30\%$ even when a choice of $1024$ out of $1200$ modes is retained. Nevertheless, as demonstrated below, this level of reconstruction accuracy is already sufficient for estimating the trend of both velocity and pressure fields.

In the following, the proposed method is applied to predict the first $1024$ temporal modes out of the total $1200$. The models $\mathbf{f}(\mathbf{s}(t))$ and $\mathbf{g}(\mathbf{s}(t))$ share the same network architecture, which is summarized in Table \ref{tab:channel-arch}. The hyperparameters for training are listed in Tab. \ref{tab:channel-hyper}. $5\%$ of the supervised set is preserved for validation.

\begin{figure}[htbp]
\centering
\includegraphics[width=1\textwidth, trim=100 0 100 0, clip]{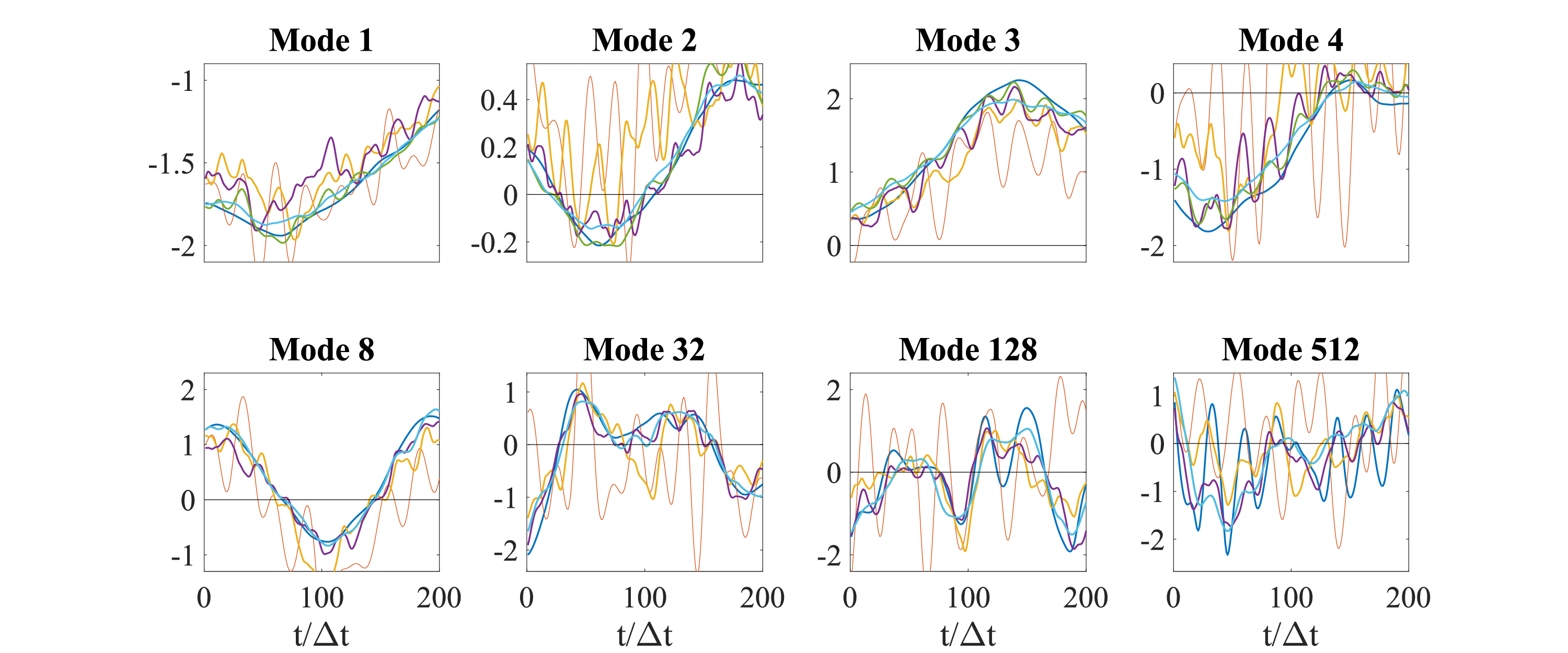}
\includegraphics[width=1\textwidth, trim=100 0 100 0, clip]{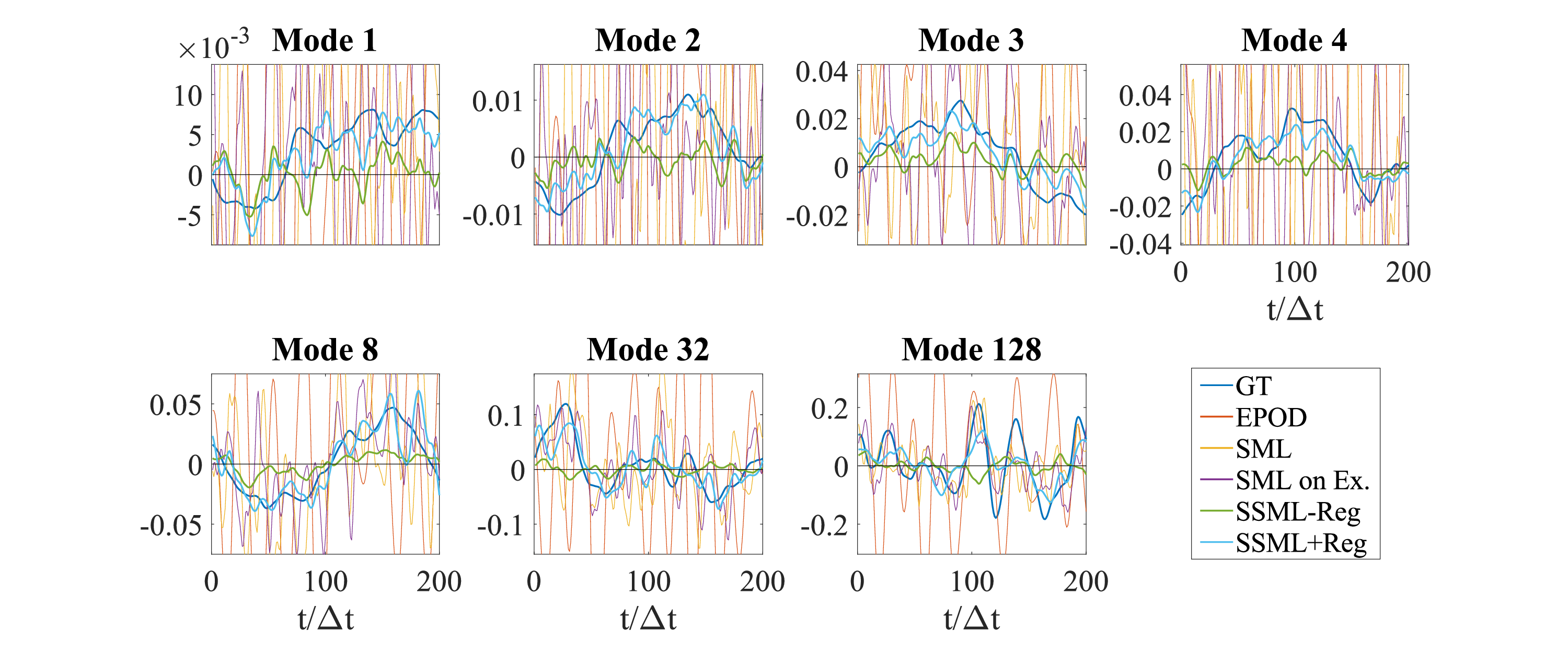}
\caption{The ground truth and predicted $\psi$ of mode 1, 2, 3, 4, 8, 32, 128, 512 (from $1^{st}$ to $2^{nd}$ row), normalized by $1/\sqrt{n_t}$. The ground truth and predicted $\psi_t$ of mode 1, 2, 3, 4, 8, 32, 128 (from $3^{rd}$ to $4^{th}$ row), normalized by $1/\sqrt{n_t}$. The black horizontal line locates the position of $0$.}\label{C_Mode}
\end{figure}

The \ac{ML} framework predicts the temporal coefficients $\mathbf{\Psi}(t)$ associated with each \ac{POD} spatial mode in Eq. \ref{eq: POD_U}, together with their temporal derivatives $\mathbf{\Psi}_t(t)$ as defined in Eq. \ref{eq: recon_ut}. Fig. \ref{C_Mode} compares the ground truth with predictions obtained using different methods. The prediction from \ac{EPOD} and the supervised \ac{ML} exhibit pronounced temporal oscillations. In particular, although a Butterworth filter has been applied on the \ac{EPOD}, as proposed in Ref.~\cite{chen2022pressure}, meaningful agreement with the ground truth is achieved only when the testing snapshots coincide with the training samples. When the supervised \ac{ML} is trained on expanded dataset, the prediction of $\mathbf{\Psi}(t)$ improves substantially. However, the temporal derivatives $\mathbf{\Psi}_t(t)$ obtained via finite differences still exhibit significant jittering. Such fluctuations are detrimental to pressure reconstruction via the Navier–Stokes equations (Eq. \ref{eq: NS}), which are highly sensitive to the temporal derivative of the velocity field. By contrast, the learning strategy that exploits the unlabelled probe data with semi-supervised \ac{ML} yields predictions that are significantly closer to the ground truth, while the $\mathbf{\Psi}_t$ estimated from $\mathbf{g}(\mathbf{s}(t))$ shows higher stability then that yielded from numerical differentiation. Moreover, after post-processing with \ac{LSM} regularization, it provides markedly improved estimates of $\mathbf{\Psi}_t(t)$, particularly for the lower-order modes.

\begin{figure}[htbp]
\centering
\includegraphics[width=1\textwidth, trim=8 0 20 0, clip]{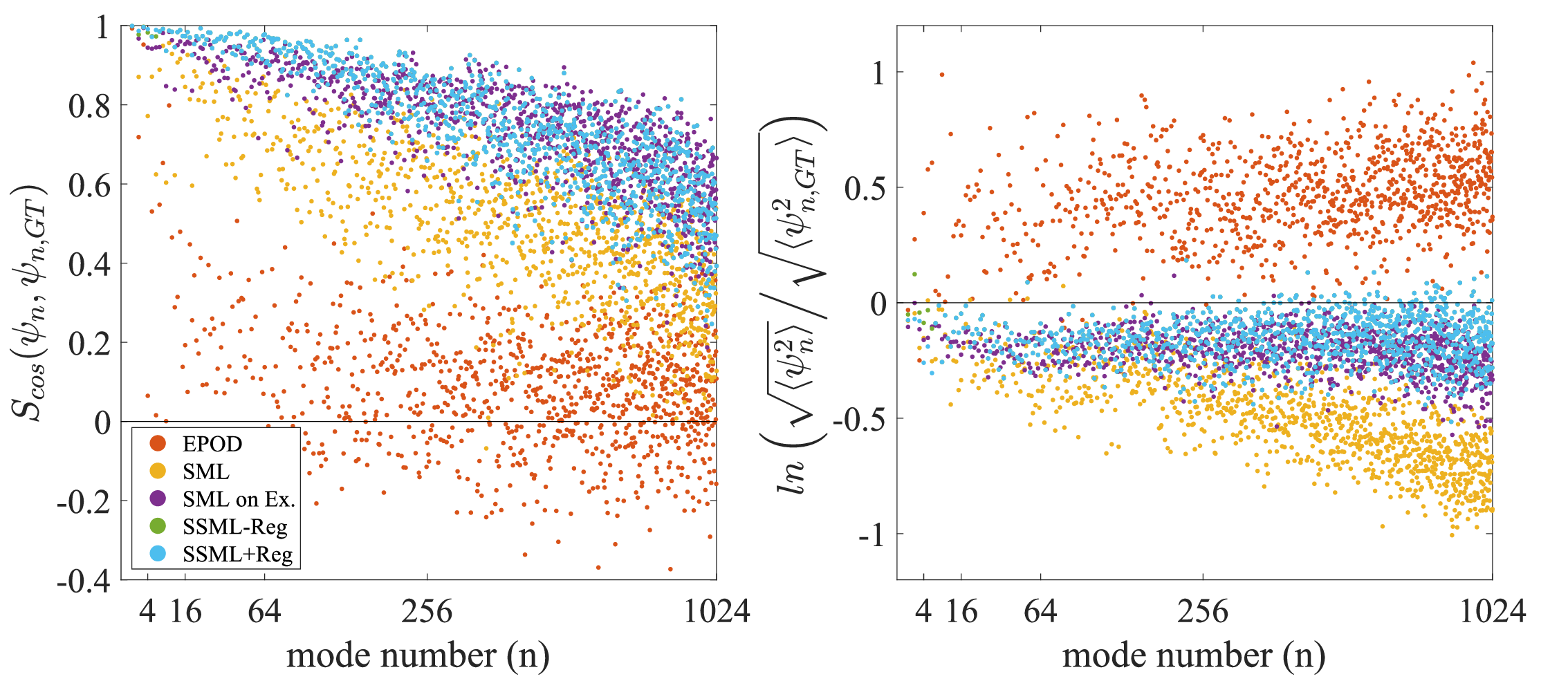}
\caption{The cosine similarity of each predicted POD temporal mode $\psi$ to the ground truth (left), the logarithmic ratio of RMS of predicted POD temporal mode $\psi$ to the counter part of ground truth (right). The horizontal black line showing the position of 0, while the horizontal axis is rescaled by the singular value $\Sigma$. Data from the simulation dataset of turbulent channel flow.}\label{C_Mode2}
\end{figure}

In addition to directly inspecting the $\psi$ curves, two quantitative metrics are employed to assess the estimation quality. The first metric is the cosine similarity between the predicted and \ac{GT} signals. Its value ranges from $-1$ to $1$, with values closer to $1$ indicating better agreement in direction, while negative values indicate counter-correlation. This metric reflects how accurately the temporal evolution is captured in a directional sense, even in the presence of overshoot or over-smooth. To complement this measure, a second metric is introduced: the ratio of \ac{RMS} value of the prediction to that of the \ac{GT}. This ratio quantifies amplitude errors and indicates whether the prediction exhibits overshoot or over-smooth.
Fig. \ref{C_Mode2} presents the results of these two metrics as a function of the mode number, where the horizontal axis is scaled by the corresponding singular values in $\mathbf{\Sigma}$. The results reveal that \ac{EPOD} yields largely random predictions beyond the $10^{th}$ mode, accompanied by significant overshoot. The supervised \ac{ML} model achieves positive cosine similarity for all modes, but exhibits the strongest over-smooth among almost all methods. The model trained on the expanded dataset improves both directional accuracy and amplitude fidelity. Finally, the model trained with semi-supervised \ac{ML} outperforms all other approaches, with additional regularization (see Sec. \ref{sect: synthesis}) further enhancing performance for the lower-order modes.

\begin{figure}[htbp]
\centering
\includegraphics[width=1\textwidth, trim=0 0 0 0, clip]{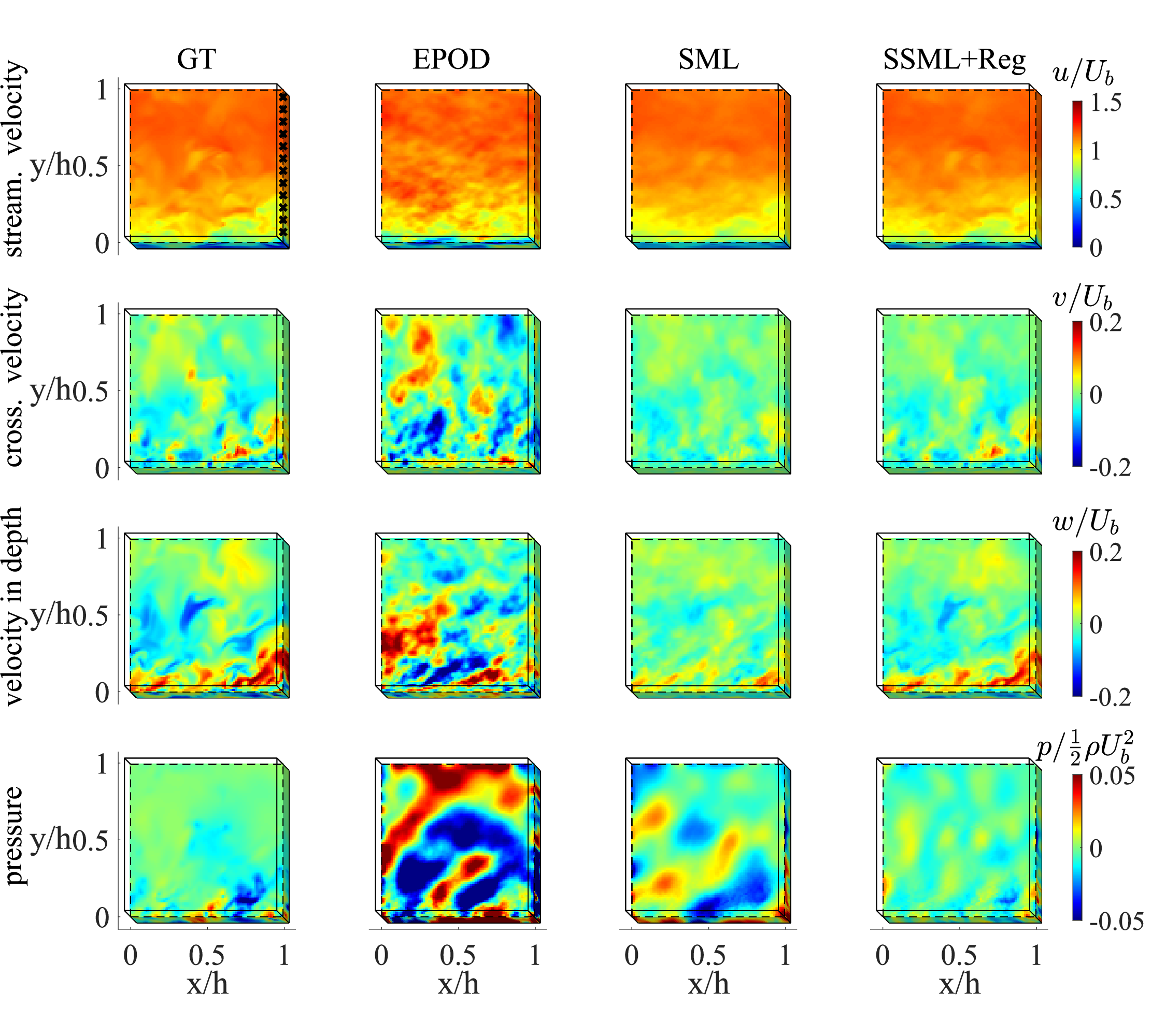}
\caption{The ground truth (the $1^{st}$ column) and predicted (from the $2^{nd}$ column) flow field from the channel data set, from top to bottom, the three components of the velocity field and pressure field. All fields are normalized, and are displayed with a vertical slice and an horizontal slice. The crossing signs on the up-left subplot stands for the position of probes.}\label{C_Field}
\end{figure}

A representative snapshot of the reconstructed flow field is shown in Fig. \ref{C_Field}. The columns correspond to the \ac{GT}, \ac{EPOD}, supervised \ac{ML}, and semi-supervised \ac{ML} after regularization, respectively. The snapshot is taken at a time instant located midway between two labelled training samples.
The velocity field estimated using \ac{EPOD} exhibits strong, spatially random motions at intermediate scales, consistent with the previously observed degradation in cosine similarity and amplitude fidelity. These inaccuracies propagate directly into the pressure computation, leading to large errors in the estimated pressure field.
Although the supervised \ac{ML} approach shows significant temporal fluctuations in the predicted $\psi$ coefficients (as seen in Fig. \ref{C_Mode}), the reconstructed velocity field primarily reflects over-smooth. As a result, the associated pressure field is also poorly estimated.
In contrast, the model trained from the proposed method produces velocity fields that closely resemble the \ac{GT}, albeit with some loss of fine-scale details. Importantly, the corresponding pressure field correctly captures both low- and high-pressure regions.
The remaining discrepancies in the semi-supervised \ac{ML} predictions can be attributed to two main factors. First, residual errors persist in the prediction of $\mathbf{\Psi}$ and $\mathbf{\Psi}_t$, as discussed above. Second, limitations arise from reconstructing a highly complex flow using a finite number of \ac{POD} modes, which is particularly significant given the high dimensionality of the present flow. Nevertheless, this work primarily aims to improve deep learning performance by more effectively utilizing both labelled and unlabelled training data; therefore, the design of more advanced flow encoders and decoders is not explored here.

\begin{table}[t]
\centering
\caption{Reconstruction errors of velocity and pressure fields obtained using different estimation methods.}
\label{tab:Ci_errors}
\begin{tabular}{l|c|ccccc}
\hline
Error & LOR & EPOD & SML & SML on Ex. & SSML-Reg & SSML+Reg \\
\hline
Velocity & 0.0198 & 0.0706 & 0.0356 & 0.0292 & 0.0282 & \textbf{0.0273} \\
Pressure & 0.0103 & 0.0609 & 0.0225 & 0.0201 & 0.0146 & \textbf{0.0127} \\
\hline
\end{tabular}
\end{table}

The estimation errors of the velocity and pressure fields for all methods are summarized in Tab. \ref{tab:Ci_errors}. For reference, the error associated with the \ac{LOR} is also included, representing the limitation imposed by the use of the current \ac{POD}-based encoder and decoder.
As shown in the table, successive methodological improvements (including the use of neural networks, training set expansion with $T_P$, semi-supervised \ac{ML}, and \ac{LSM} regularization) lead to consistent reductions in estimation error. The final prediction achieves accuracy that is very close to the physical lower bound bounded by the \ac{LOR}.

\begin{figure}[htbp]
\centering
\includegraphics[width=1\textwidth, trim=0 0 0 0, clip]{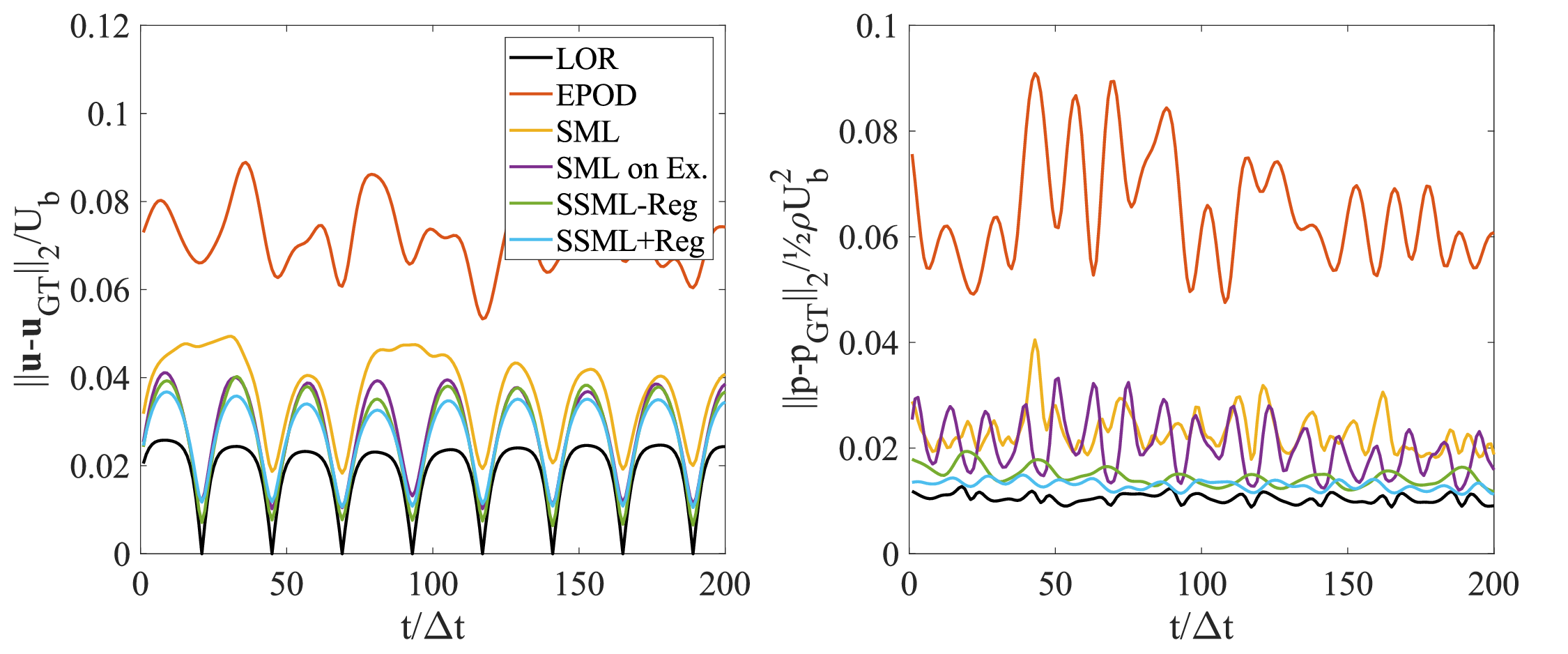}
\caption{Turbulent channel flow dataset. Temporal evolution of reconstruction errors for the velocity field (left) and pressure field (right).}\label{fig: Ci_time}
\end{figure}

Fig. \ref{fig: Ci_time} shows the time evolution of the reconstruction errors for both velocity and pressure fields. The velocity reconstruction error exhibits a periodic pattern with a period of $24$ frames, which coincides with the sampling interval of the labelled training snapshots extracted from the testing dataset. However, this periodicity is mainly associated with the intrinsic limitation of the \ac{POD} representation that the spatial basis trained from limited dataset can not cover all the situations, whereas the role of errors in the estimation of $\mathbf{\Psi}$ remains less clear and is discussed in relation to Fig. \ref{C_Mode}.
Overall, the velocity reconstruction error is progressively reduced as the proposed methods are applied, with the semi-supervised \ac{ML} achieving the best performance, particularly for time instants located between labelled training samples.

The pressure reconstruction error exhibits oscillations at approximately twice the frequency of the velocity-field error. This behaviour can be interpreted as a consequence of the temporal derivative: near the peaks and troughs of the velocity field error, the predicted temporal coefficients vary more slowly, thus introduce less error to the pressure gradients. 
This phenomenon is more pronounced in supervised \ac{ML} on expanded training dataset, where the temporal derivative of the velocity field is obtained less consistently.
After applying the semi-supervised \ac{ML} strategy, these oscillations are effectively suppressed and the error curves become significantly flatter. In most time frames, the \ac{LSM} regularization further improves the pressure estimation accuracy.

\begin{figure}[htbp]
\centering
\includegraphics[width=1\textwidth, trim=0 0 0 0, clip]{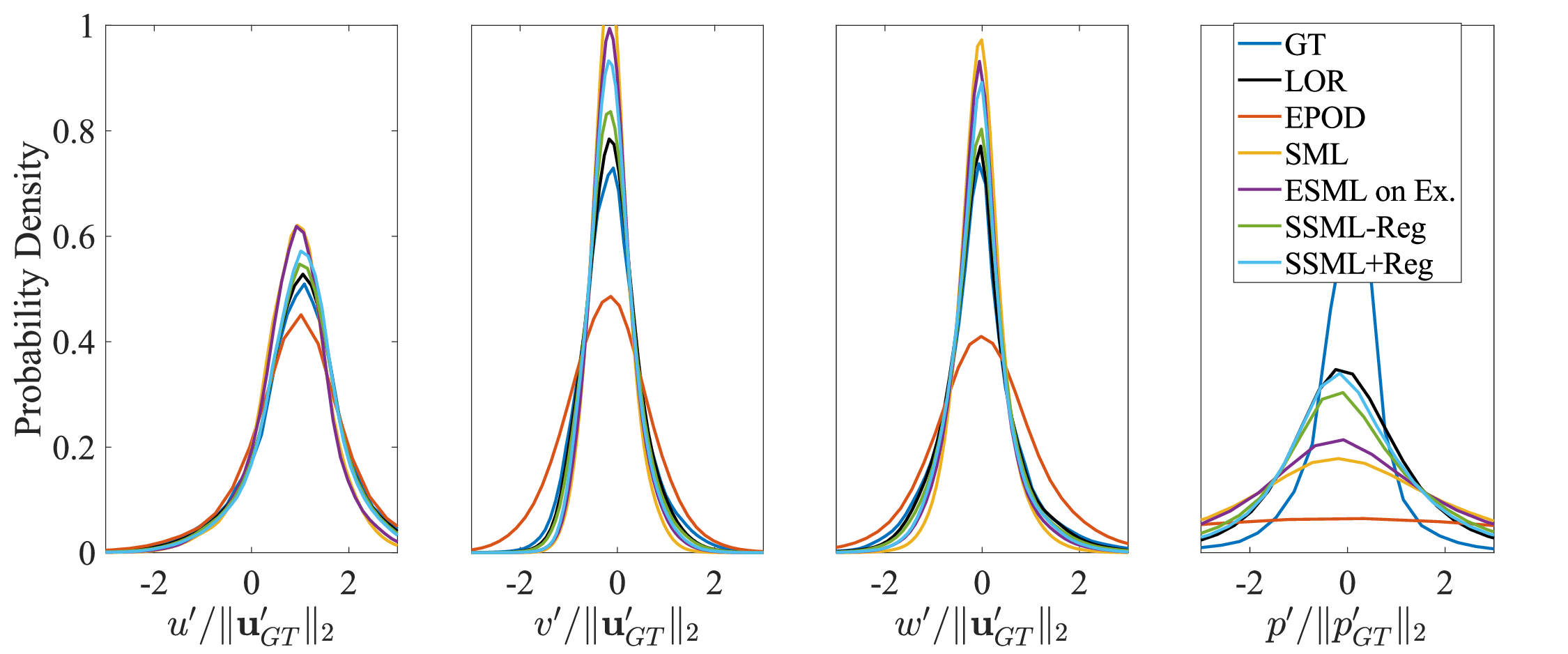}
\caption{PDFs of fluctuating streamwise velocity and pressure for ground truth, LOR, and all reconstructions, normalized by RMS of $\mathbf{u}'_{GT}$ or $p'_{GT}$ and evaluated over all testing frames of the turbulent channel flow dataset.}\label{fig: Ci_PDF}
\end{figure}

\begin{figure}[htbp]
\centering
\includegraphics[width=1\textwidth, trim=0 0 0 0, clip]{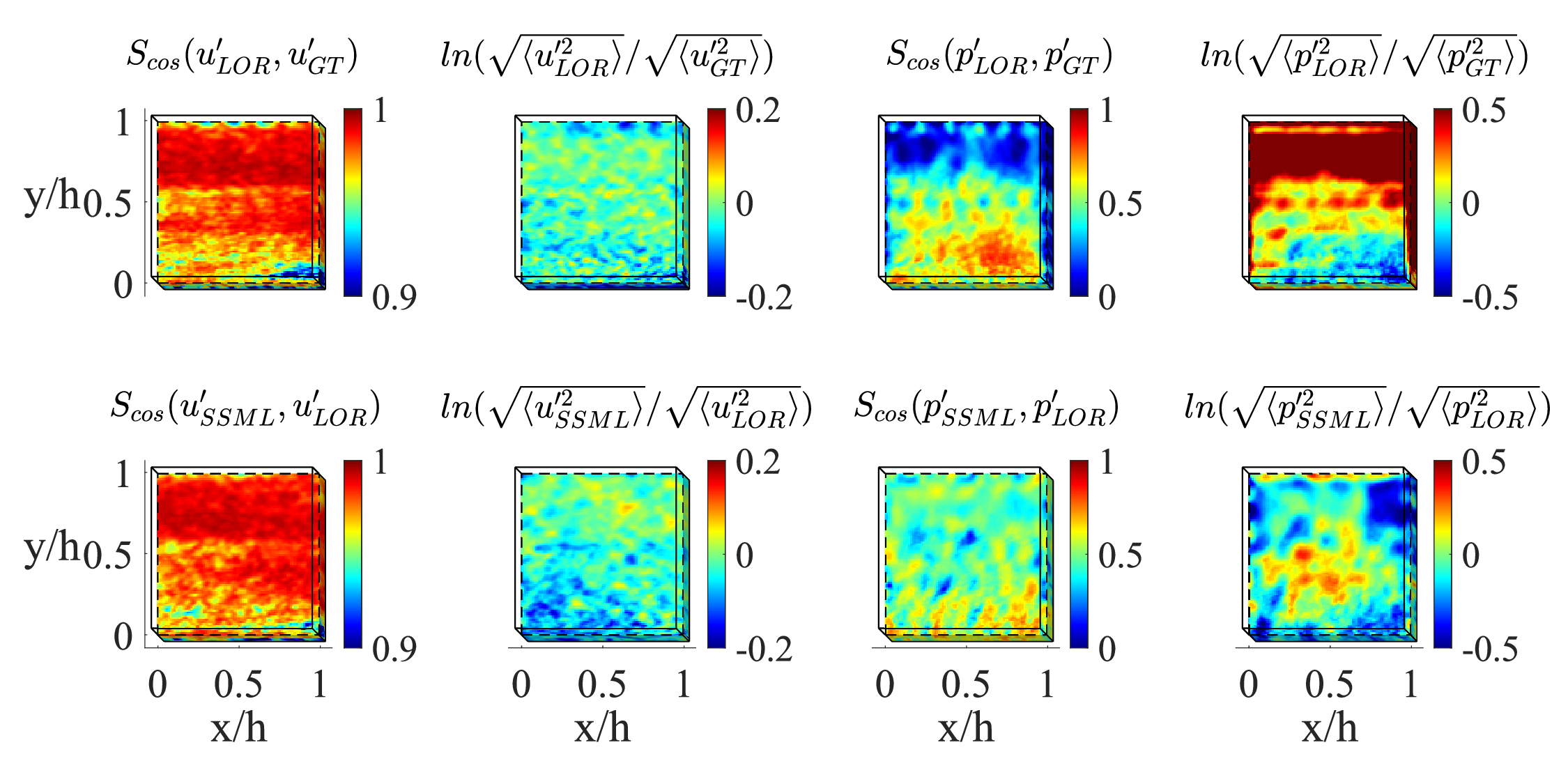}
\caption{Cosine similarity (columns 1, 3) and logarithmic RMS fluctuation ratio (columns 2, 4) of the temporally averaged streamwise velocity (columns 1–2) and pressure (columns 3–4) fields. Top: LOR versus ground truth. Bottom: semi-supervised \ac{ML}, with regularization, reconstruction trained on unlabelled samples versus LOR.}\label{fig: Ci_LOR_map}
\end{figure}

Figure \ref{fig: Ci_PDF} compares the \acp{PDF} of the \ac{GT} and the estimated fields. Consistent with the single-frame results, \ac{EPOD} produces excessive spurious motions, leading to a much broader distribution than the \ac{GT}. In contrast, the progression from supervised \ac{ML} to semi-supervised \ac{ML} yields progressively improved distributions that move closer to the \ac{GT}.
After \ac{LSM} regularization, the velocity-field \ac{PDF} shifts slightly away from the \ac{GT}, while the pressure-field \ac{PDF} is further improved and approaches that of the \ac{LOR}, which represents the performance limit of the current encoder.
The limitations of the \ac{POD}-based representation are further illustrated in Fig. \ref{fig: Ci_LOR_map}. The \ac{LOR} reconstructs the central-region motions reasonably well but exhibits errors near the wall, which propagate throughout the domain during pressure integration. A similar level of near-wall velocity error is observed in the semi-supervised \ac{ML}; however, the resulting pressure errors are substantially reduced.

\begin{figure}[htbp]
\centering
\includegraphics[width=1\textwidth, trim=0 0 0 0, clip]{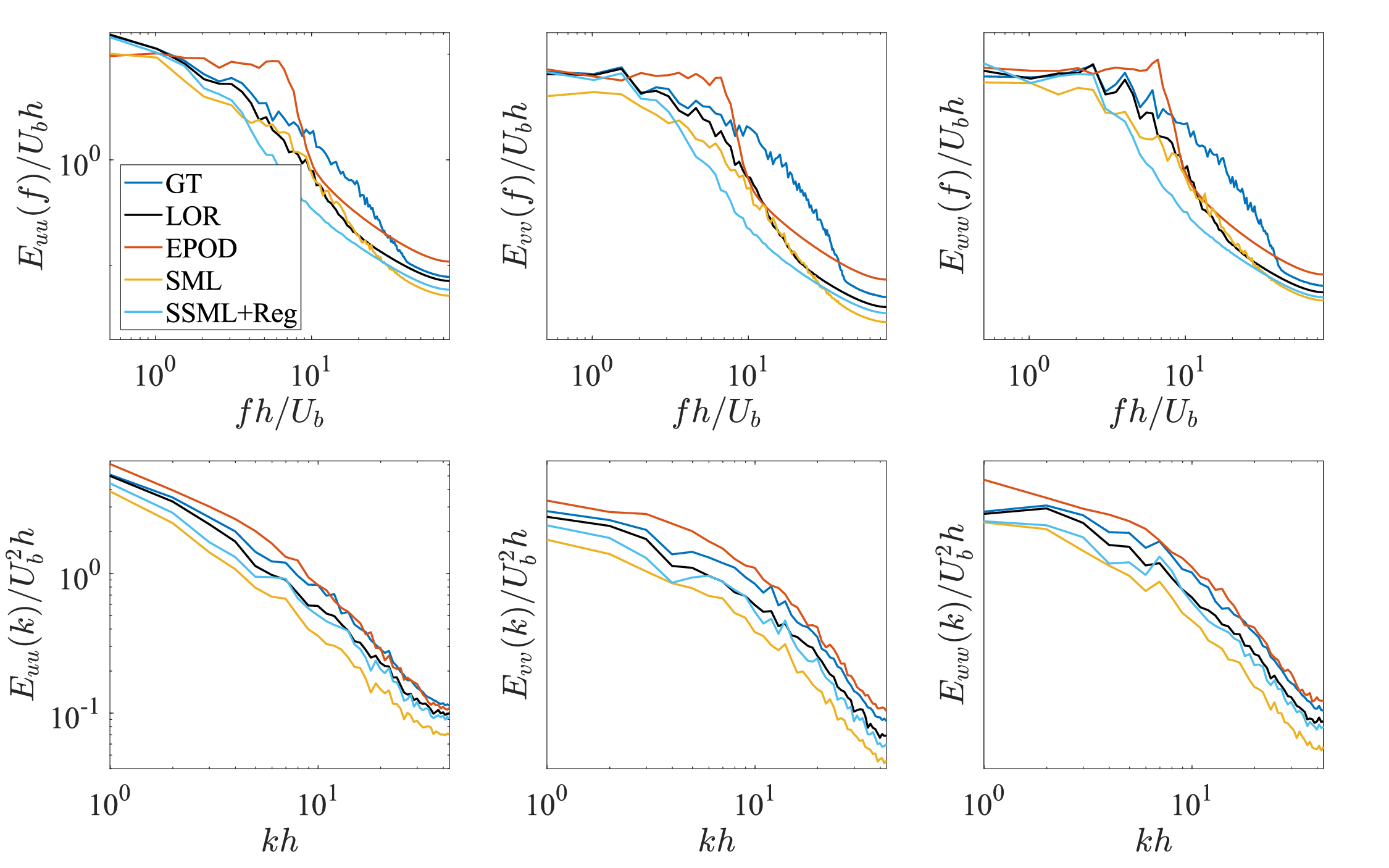}
\caption{Power spectra of the velocity field for GT, \ac{LOR}, and all reconstruction techniques. Top: temporal frequency spectra; bottom: wavenumber spectra. Columns show the three velocity components.}\label{fig: Ci_FFT}
\end{figure}

Figure \ref{fig: Ci_FFT} presents the power spectra of the three velocity components for the turbulent channel-flow dataset, performed at $y/h = 0.11$. For all components, the \ac{LOR} fails to recover mid-to-high-frequency motions, particularly around the effective sampling frequency of the training data, and exhibits slightly attenuation across all scales.
The \ac{EPOD} reconstruction introduces broadband noise relative to the \ac{LOR}. In particular, the spectral energy at low frequencies and large scales is even higher than that of the ground truth, indicating the presence of spurious motions.
The supervised \ac{ML} suppresses this noise but attenuates energy across the full spectrum. In contrast, the semi-supervised \ac{ML} model shows reduced energy only in the mid-frequency range, while achieving better recovery at low and high frequencies, closely matching \ac{LOR}. When examined in wavenumber space, the semi-supervised \ac{ML} exhibits higher consistent spectral energy across all wavenumbers, indicating improved spatial reconstruction. This suggests that the attenuated mid-frequency response observed in the temporal spectra may not correspond to missing physical motions, but rather to the suppression of spurious fluctuations.

\section{Experimental application}
\label{sec: experimental}

An experimental validation case is carried out on \ac{PIV} measurements in a water tunnel on a two-dimensional wing model with a NACA 0018 airfoil section. Detailed information is provided in Ref.~ \cite{chen2022pressure}. The wing is placed at an angle of attack of $10^\circ$. The Reynolds number based on the chord length ($80 mm$) and the freestream speed ($0.06 m/s$) is $4800$. Time-resolved two-dimensional \ac{PIV} measurements are acquired at a sampling frequency of $30 \mathrm{Hz}$. To reduce measurement noise, the \ac{AMIC} method \cite{chen2025advection} is applied. A subdomain of $110 \times 70$ vectors with a uniform spacing of $1.20\,\mathrm{mm}$ is considered.

Eleven points at the downstream end of the domain are considered to simulate velocity probes for the streamwise component. As in the synthetic case, the probe signals are extracted from the high-temporal-resolution \ac{PIV} velocity field. For each probe, a temporal sequence of $100$ samples is collected, mimicking the convection of flow information across the field. The training dataset consists of $1200$ labelled snapshots randomly downsampled from the time-resolved \ac{PIV} measurements, together with unlabelled snapshots that are $14$ times more numerous than the labelled ones. The testing dataset comprises $500$ consecutive frames. Both networks $\mathbf{f}(\mathbf{s}(t))$ and $\mathbf{g}(\mathbf{s}(t))$ adopt the same model architecture listed in Tab. \ref{tab:wing-arch}, and are used to predict the first $1024/1200$ temporal \ac{POD} modes. $5\%$ of the training set are kept for validation while the hyper-parameters are listed in Tab. \ref{tab:wing-hyper}.

\begin{figure}[htbp]
\centering
\includegraphics[width=1\textwidth, trim=0 0 30 0, clip]{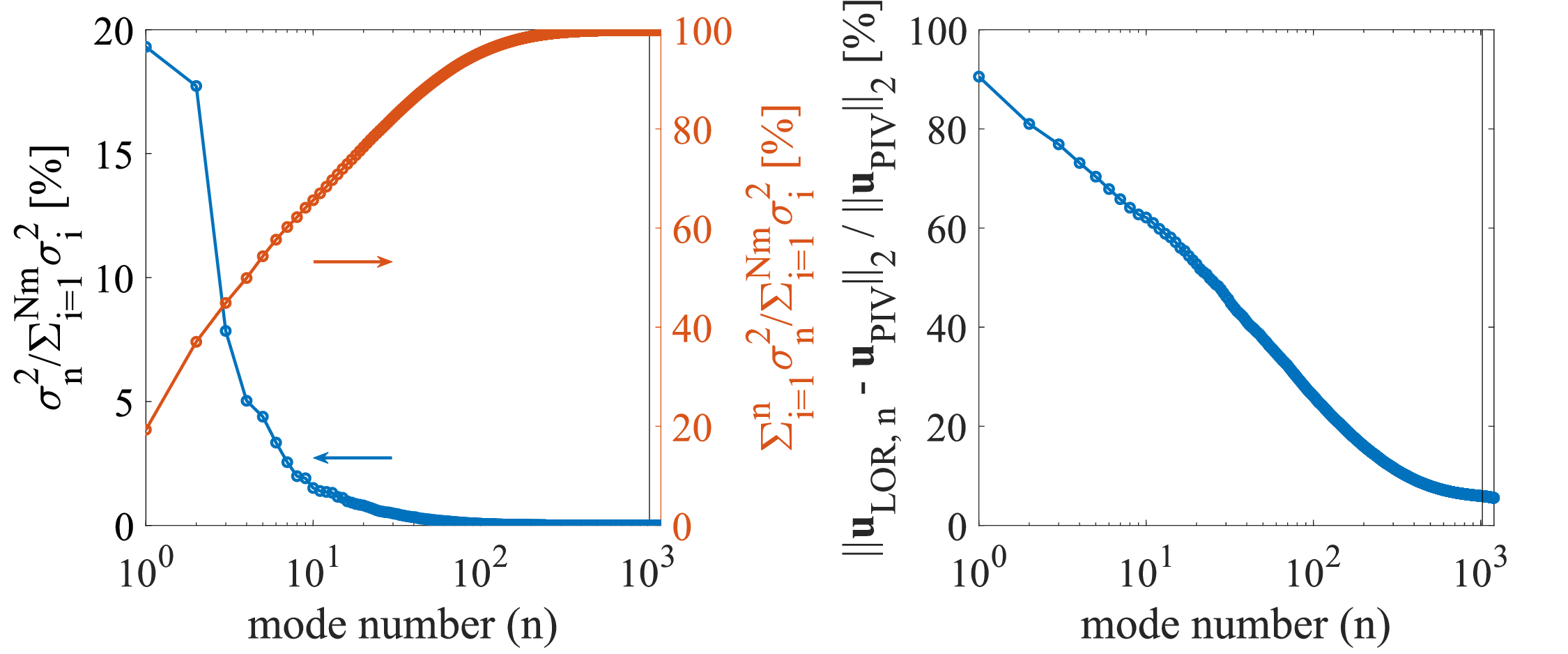}
\caption{POD eigenspectrum and cumulative energy versus mode number (left) for the wing wake dataset. Relative RMS error of LOR (right).}\label{fig: W_POD}
\end{figure}

\begin{figure}[htbp]
\centering
\includegraphics[width=1\textwidth, trim=100 0 100 0, clip]{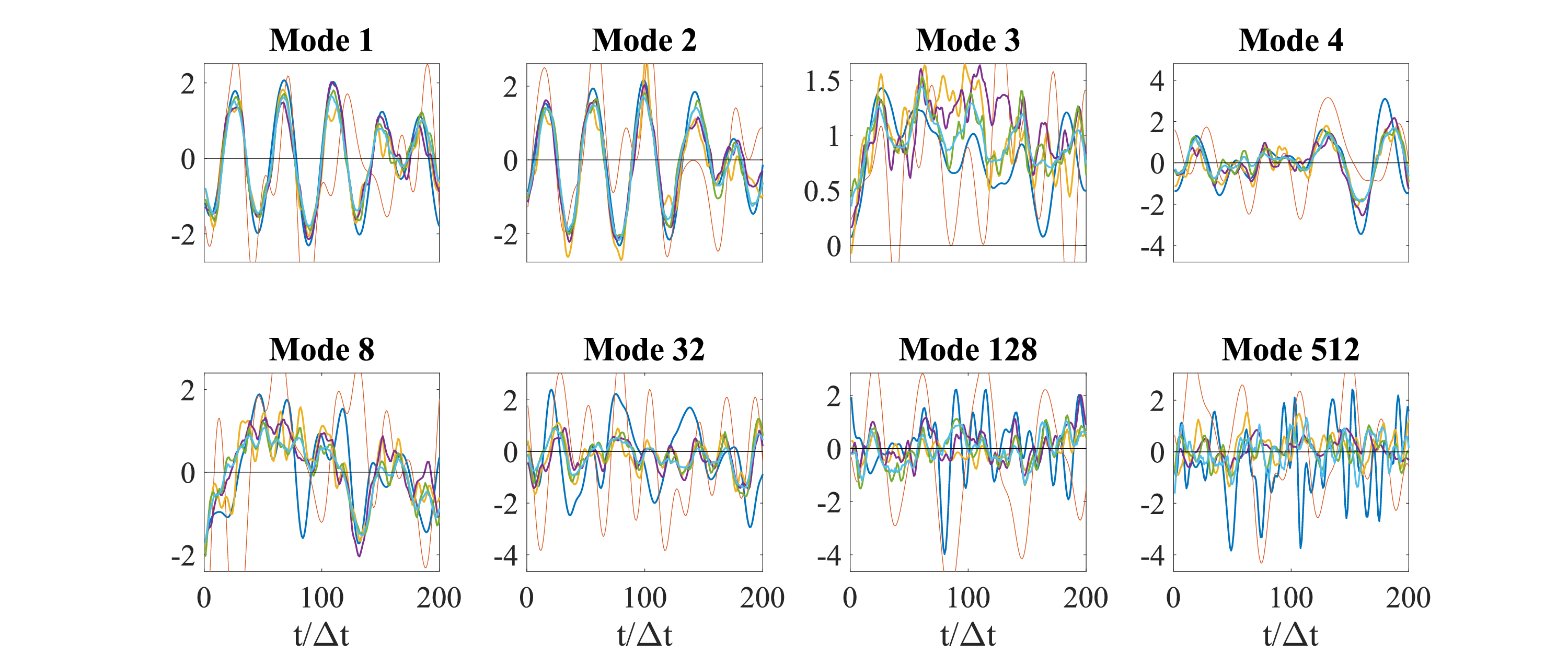}
\includegraphics[width=1\textwidth, trim=100 0 100 0, clip]{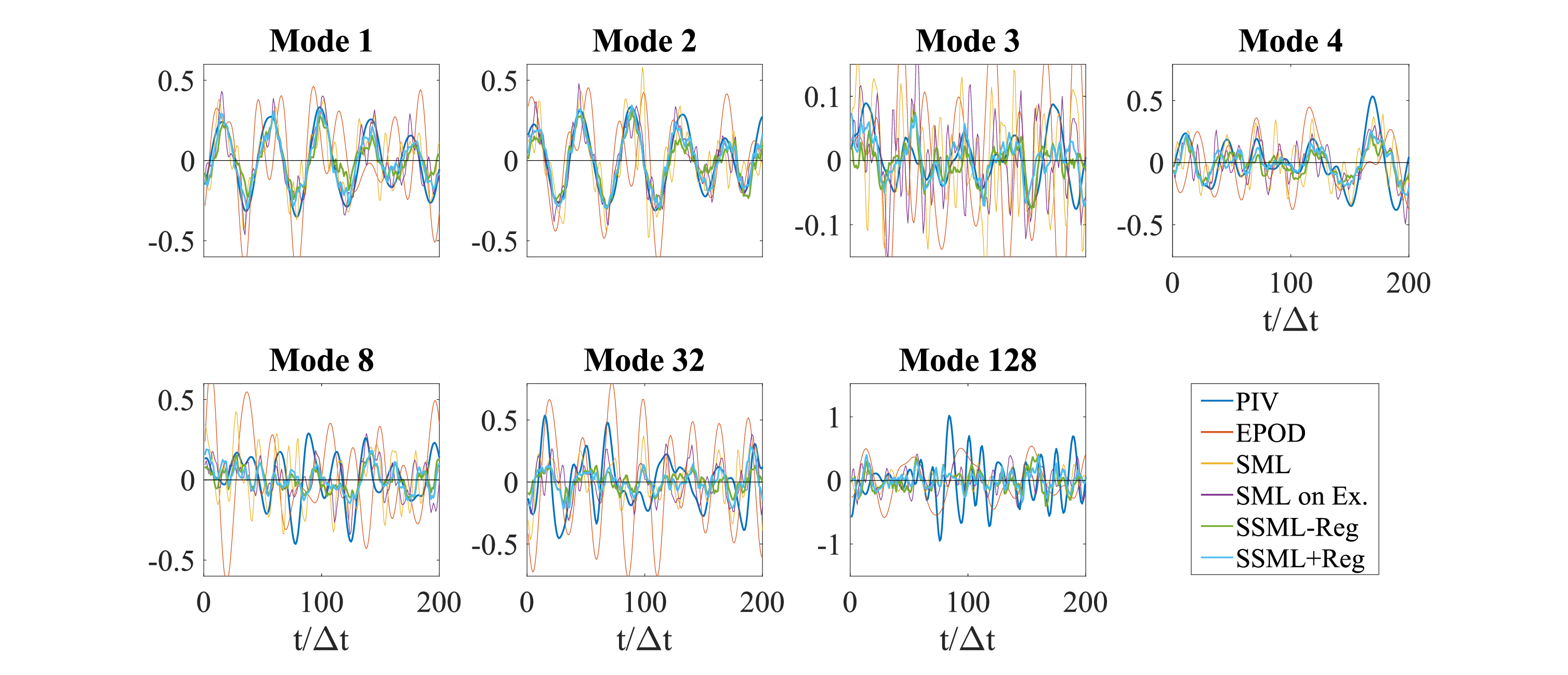}
\caption{ Comparison of \ac{PIV} reference and predicted spatial modes $\psi$ (top two rows) and temporal modes $\psi_t$ (bottom two rows) for selected mode numbers for the wing wake dataset. All modes are normalized by $1/\sqrt{n_t}$, and the black horizontal line marks zero.}\label{fig: W_Mode}
\end{figure}

The \ac{POD} squared singular values shown in Fig. \ref{fig: W_POD} indicates that the flow in this case is significantly less complex than the synthetic turbulent channel flow, and that the \ac{LOR} is able to recover the dominant flow features with high fidelity.
Fig. \ref{fig: W_Mode} compares the \ac{PIV} result with predictions from different methods for both the temporal modes $\mathbf{\Psi}$ and their temporal derivatives $\mathbf{\Psi}_t$ in the wing wake dataset.
All \ac{ML} methods reconstruct the temporal coefficients $\mathbf{\Psi}$ accurately for modes 1, 2, and 4, whereas \ac{EPOD} exhibits relatively large deviations even after filtering. Mode 3 is notably aperiodic, contaminated by the pump motion of the water tunnel at low flow speeds; here, the semi-supervised \ac{ML} outperforms both \ac{EPOD} and supervised \ac{ML}. For higher-order modes (from approximately mode 128 onward), none of the deep-learning approaches fully reproduces the amplitude of the fluctuations observed in the \ac{PIV} data. Nevertheless, in predicting the temporal derivatives $\mathbf{\Psi}_t$, the proposed method consistently outperforms the alternatives. A subsequent \ac{LSM} regularization further smooths the predicted curves and improves temporal consistency.

\begin{table}[htp]
  \centering
  \caption{The model structure of the airfoil wake dataset}
  \label{tab:wing-arch}
  \begin{tabular}{lcccc}
    \hline
    \textbf{Layer \#} & \textbf{Type}       & \textbf{Output Units} & \textbf{Activation} & \textbf{Dropout} \\
    \hline
    0 & Input           & $100\times 11$ & ---   & ---    \\
    1 & Dense           & 2048           & tanh  & 10\%   \\
    2 & Dense           & 2048           & tanh  & 10\%   \\
    3 & Dense           & 1024           & tanh  & 10\%   \\
    4 & Dense           & 1024           & tanh  & 10\%   \\
    5 & Output Dense    & 1024           & linear& ---    \\
    \hline
  \end{tabular}
\end{table}

\begin{table}[htp]
\centering
\begin{tabular}{c c c c}
\hline
hyperparameter   & Stage 1          & Stage 2          & Stage 3 \\
\hline
$C_{11}$         & $0$              & --               & --               \\
$C_{21}$         & --               & $0$              & --               \\
$C_{31}$         & --               & --               & $1\times10^{-12}$ \\
$C_{32}$         & --               & --               & $0$              \\
optimizer        & Adam             & Adam             & Adam             \\
learning rate    & $1\times10^{-4}$ & $1\times10^{-4}$ & $5\times10^{-6}$ \\
batch size       & $128$            & $128$            & $128$            \\
number of epochs & $800$            & $800$            & $400$            \\
LP               & $3$              & $3$              & $3$              \\
CP               & $4$              & $6$              & $6$              \\
CU               & --               & $0.2$            & $0.2$            \\
\hline
\end{tabular}
\caption{Hyperparameters used in the wing case.}
\label{tab:wing-hyper}
\end{table}

\begin{figure}[htbp]
\centering
\includegraphics[width=1\textwidth, trim=8 0 20 0, clip]{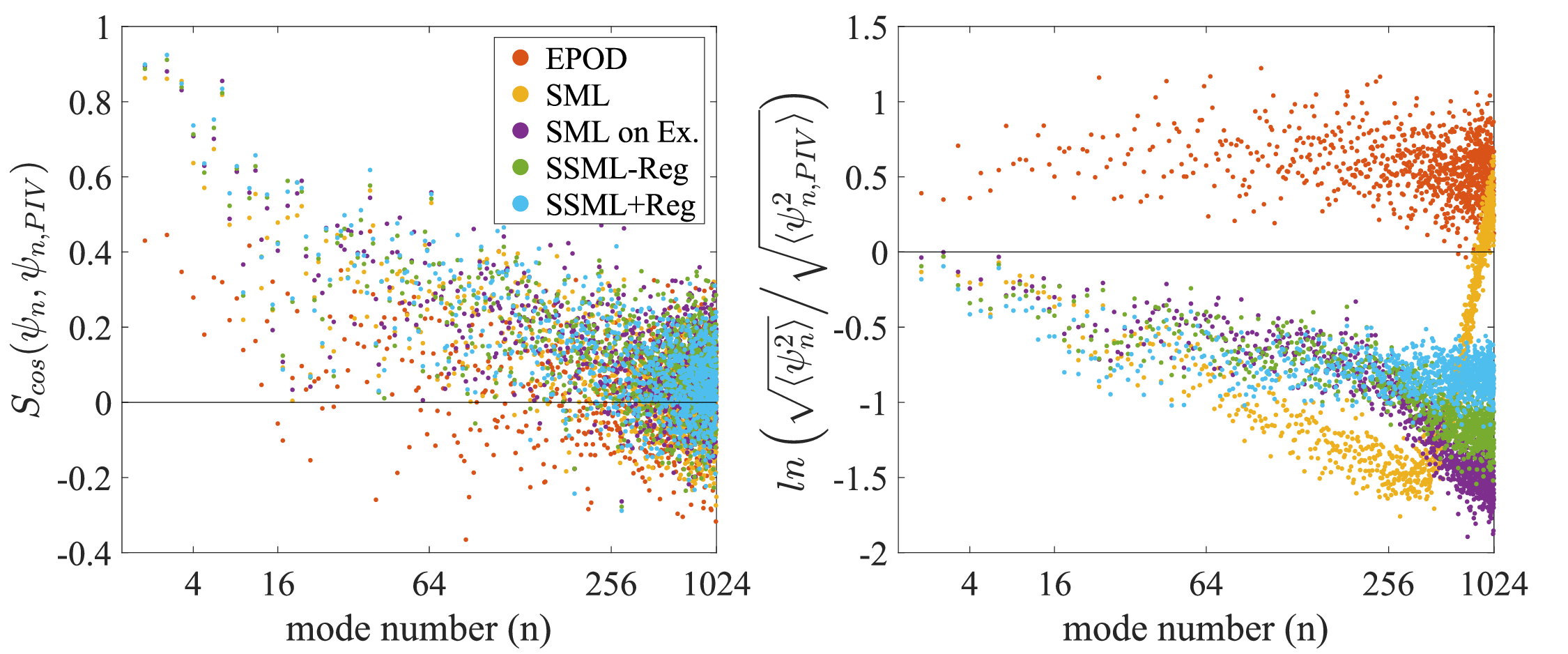}
\caption{Cosine similarity of predicted POD temporal modes $\psi$ with respect to the \ac{PIV} field (left) and their logarithmic RMS ratio relative to the \ac{PIV} field (right) for the wing wake dataset. The horizontal black line denotes zero, and the horizontal axis is rescaled by the singular value $\Sigma$.}\label{fig: W_ModeII}
\end{figure}

The cosine similarity and logarithmic \ac{RMS} ratio shown in Fig. \ref{fig: W_ModeII} indicate that the prediction accuracy degrades progressively from low to high modes. Nevertheless, retaining modes up to approximately mode $256$ remains beneficial for velocity reconstruction. The semi-supervised \ac{ML} delivers the best overall performance across most modes. On the observation on the amplitude of estimated $\mathbf{\Psi}$, \ac{EPOD} tends to produce overshooting estimations, whereas the supervised \ac{ML} generally yields over-smooth predictions up to about mode 512, followed by overshoot in the highest-order modes. Compared with the model trained on the expanded dataset, the semi-supervised \ac{ML} exhibits stronger over-smooth. This behaviour may be related to the unsupervised training constraint, which encourages similar predictions from neighbouring probe signals and may therefore suppress rapid small-scale temporal variations. The \ac{LSM} regularization further smooths the predicted temporal coefficients.

\begin{figure}[htbp]
\centering
\includegraphics[width=1\textwidth, trim=40 0 18 0, clip]{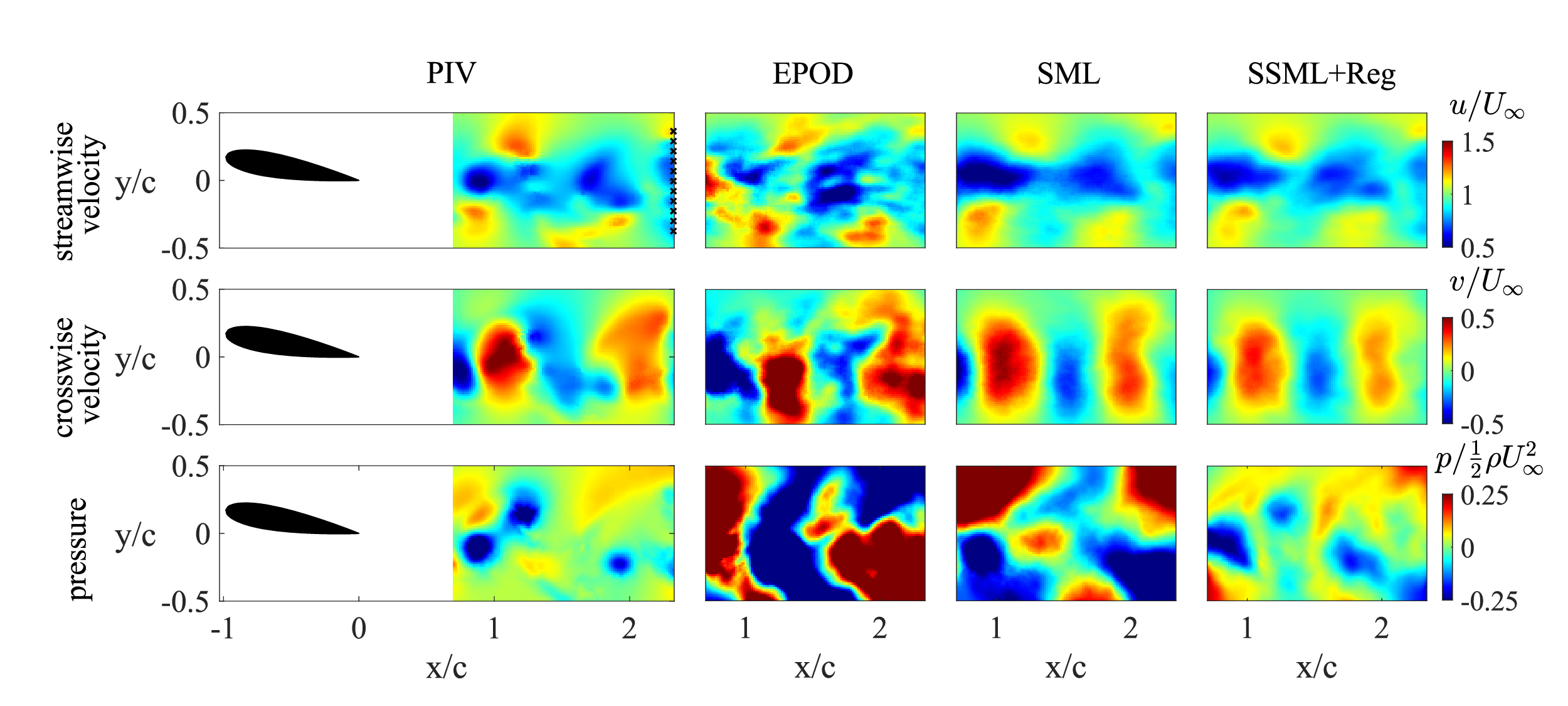}
\caption{Comparison of ground-truth (first column) and predicted (remaining columns) normalized flow fields for the wing wake dataset. Rows correspond to the streamwise velocity, crosswise velocity, and pressure fields. The probe locations are indicated by cross symbols in the upper-left subplot.}\label{fig: W_F}
\end{figure}

Fig. \ref{fig: W_F} presents an arbitrary snapshot of the reconstructed velocity and pressure fields, together with the corresponding \ac{PIV} reference. Constrained by the limited training dataset, \ac{EPOD} fails to provide a reliable reconstruction. Both the supervised \ac{ML} and semi-supervised \ac{ML} recover the dominant large-scale flow structures, although fine-scale details are not preserved.
The inter-frame evolution predicted by semi-supervised \ac{ML} appears smoother and more consistent than that of the supervised counterpart (not shown in the figure). As a consequence, the reconstructed pressure field is closer to the \ac{PIV} reference, and spurious high-magnitude values near the domain corners are substantially reduced.

\begin{figure}[htbp]
\centering
\includegraphics[width=1\textwidth, trim=0 0 0 0, clip]{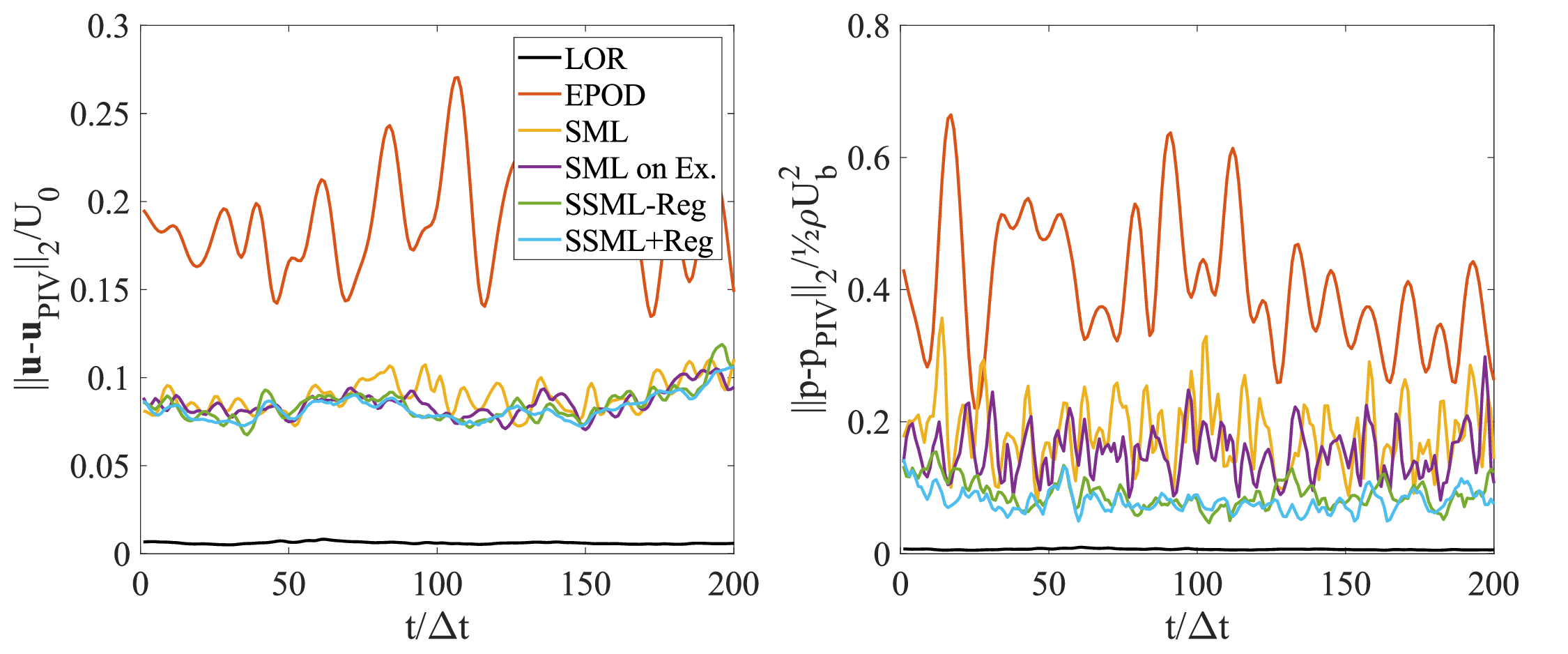}
\caption{Temporal evolution of reconstruction errors for the velocity field (left) and pressure field (right) for the wing wake dataset.}\label{fig: W_err_t}
\end{figure}

The time evolution of the velocity and pressure reconstruction errors is shown in Fig. \ref{fig: W_err_t}, and the corresponding averaged values are summarized in Tab. \ref{tab:W_errors}. \ac{EPOD} exhibits relatively large errors in the velocity field reconstruction. \ac{ML}-based methods substantially reduce these errors, and the proposed supervised \ac{ML} on expanded dataset provides a further improvement. Finally, the application of semi-supervised \ac{ML} leads to a more noticeable enhancement especially in pressure field estimation accuracy.

\begin{table}[t]
\centering
\label{tab:W_errors}
\begin{tabular}{l|c|ccccc}
\hline
Error & LOR & EPOD & SML & SML on Ex. & SSML-Reg & SSML+Reg \\
\hline
Velocity & 0.00687 & 0.2198 & 0.0969 & 0.0948 & 0.0931 & \textbf{0.0907} \\
Pressure & 0.00631 & 0.2922 & 0.1185 & 0.1183 & 0.0672 & \textbf{0.0519} \\
\hline
\end{tabular}
\end{table}

\begin{figure}[htbp]
\centering
\includegraphics[width=1\textwidth, trim=0 0 0 0, clip]{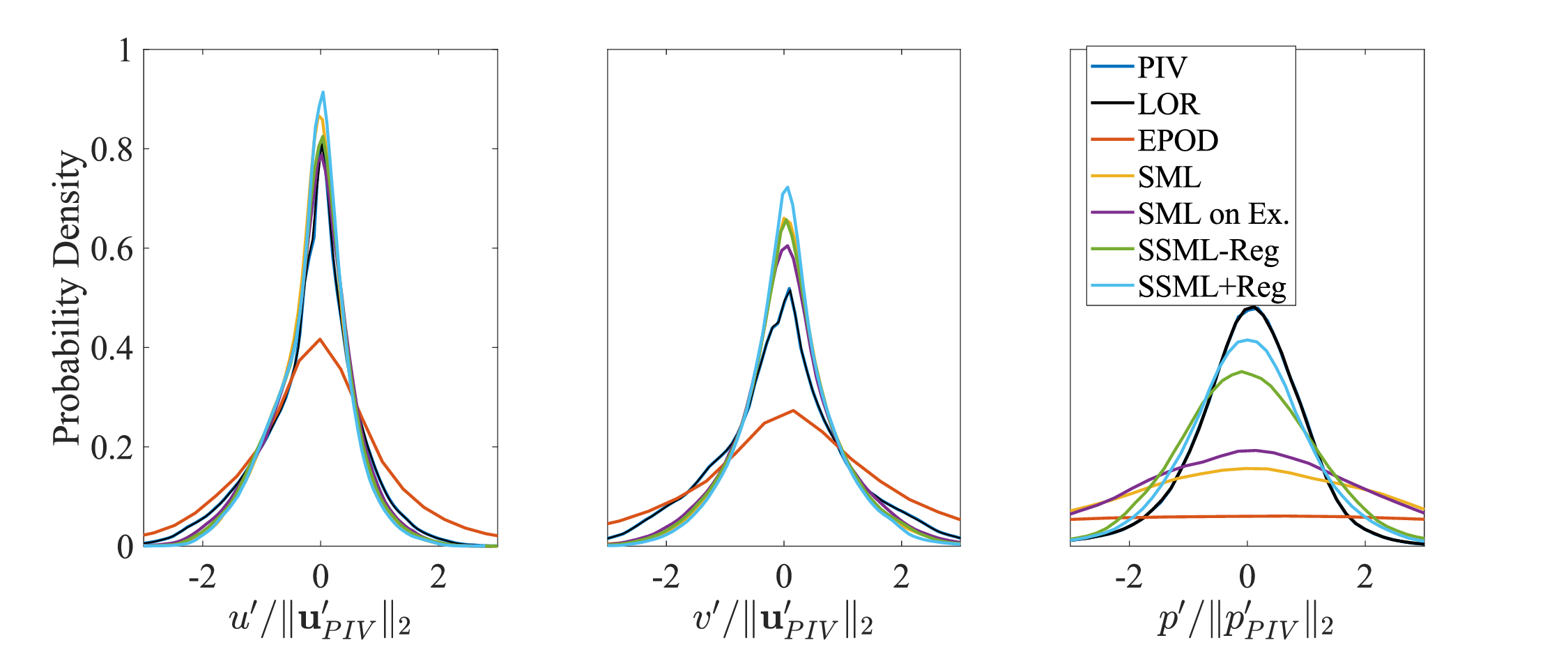}
\caption{Probability density functions of fluctuating velocity and pressure for the original \ac{PIV} fields, \ac{LOR}, and reconstructed fields for the wing wake dataset. All PDFs are computed over the testing set and normalized by the RMS of the fluctuations, $\mathbf{u}'_{PIV}$ and $p'_{PIV}$.}\label{fig: W_PDF}
\end{figure}

\begin{figure}[htbp]
\centering
\includegraphics[width=1\textwidth, trim=0 0 0 0, clip]{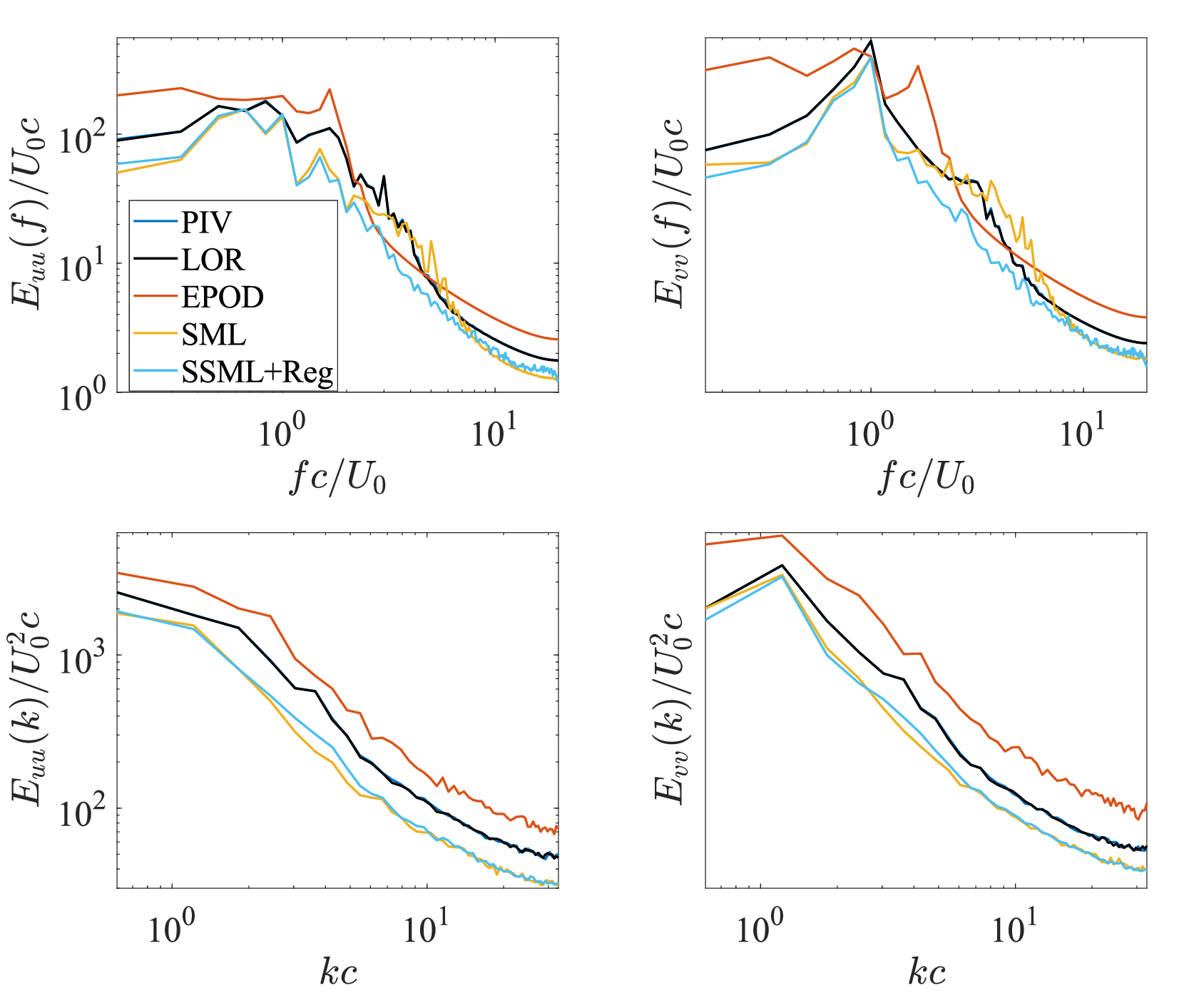}
\caption{Temporal-frequency (top) and wavenumber (bottom) power spectra of the two velocity components for the \ac{PIV} fields, \ac{LOR}, and reconstructed fields for the wing wake dataset.}\label{fig: W_FFT}
\end{figure}

The probability density functions of the velocity fluctuations $u'$, $v'$ and the pressure fluctuation $p'$ are shown in Fig. \ref{fig: W_PDF}, and the power spectra of $u'$ and $v'$ are presented in Fig. \ref{fig: W_FFT}. The \ac{LOR} reproduces the \ac{PIV} distributions closely for both velocity and pressure. For the velocity field, all \ac{ML}-based methods yield narrower \ac{PDF}s than \ac{PIV}, indicating reduced variance. The proposed semi-supervised \ac{ML} shifting the distributions closer to the \ac{PIV} reference compared with supervised \ac{ML}. For the pressure field, the proposed semi-supervised \ac{ML} produces a more concentrated distribution with improved agreement with \ac{PIV}.
The power spectra show that \ac{EPOD} introduces excess energy over a broad range of frequencies and wavenumbers, whereas the \ac{ML}-based methods exhibit attenuated spectral energy. Compared with supervised \ac{ML}, the use of semi-supervised \ac{ML} improves the recovery of low-frequency content while maintaining reduced energy at higher frequencies.

\section{Conclusion}
\label{sec: conclusion}

This work proposes a semi-supervised \ac{ML} framework for time-resolved flow field reconstruction from sparse probe measurements, with a particular focus on improving the utilization of unlabelled data. Two neural networks are trained to predict the temporal coefficients of \ac{POD} modes and their temporal derivatives, respectively, and a regularization step based on \ac{LSM} is introduced to enforce temporal consistency between the two predictions.

The method is validated on both a synthetic turbulent channel flow dataset and an experimental airfoil wake dataset. In the turbulent channel flow case, the proposed framework significantly improves the prediction of temporal coefficients compared with \ac{EPOD} and purely supervised \ac{ML}, especially for intermediate and higher-order modes. The incorporation of unlabelled probe data via semi-supervised \ac{ML} reduces temporal jitter and improves the consistency of predicted temporal derivatives, which is critical for pressure field estimation. As a result, both velocity and pressure reconstruction errors are reduced, with substantially larger relative gains observed in the pressure field due to its strong sensitivity to temporal inconsistencies. This highlights that the primary benefit of the proposed approach lies in enhancing pressure reconstruction accuracy rather than solely improving velocity predictions, approaching the accuracy limit imposed by the low-order \ac{POD} representation.

In the experimental wing wake case, where the flow exhibits lower dimensionality, the method demonstrates stable and robust performance despite the limited amount of labelled \ac{PIV} data. The use of unlabelled probe data leads to improved temporal smoothness and more reliable pressure reconstruction, while maintaining accurate recovery of dominant flow structures. Notably, the improvements in pressure estimation are more pronounced than those in the velocity field, owing to the reduction of spurious temporal fluctuations that directly affect pressure gradients. Spectral and statistical analyses further confirm that the proposed approach suppresses spurious high-frequency content and improves agreement with experimental measurements.

Overall, the results demonstrate that enhancing the exploitation of unlabelled probe data is an effective strategy for improving data-driven flow reconstruction, particularly when accurate temporal derivatives are required. The method is applicable to advection-dominated flows, in which advected features are sampled by the probes and the training set can be expanded by the convective model. Importantly, the framework yields its greatest impact on pressure field reconstruction, enabling more physically consistent and robust pressure estimates compared to conventional approaches. The proposed framework provides a practical and scalable solution for probe-based flow estimation and pressure reconstruction in situations where time-resolved velocity measurements are limited.

\section*{Acknowledgements}
This project has received funding from the European Research Council (ERC) under the European Union’s Horizon 2020 research and innovation programme (grant agreement No 949085).

\section*{Code availability}
The code supporting the findings of this study is openly available at \url{https://github.com/woiiiiow/semi_supervised_ML}.





\bibliographystyle{elsarticle-num} 
\bibliography{BIB}






\end{document}